# 130 years of spectroheliograms at Paris-Meudon observatories (1892-2022)

J.-M. Malherbe, Observatoire de Paris, PSL Research University, CNRS, LESIA, Meudon, France

Email : Jean-Marie.Malherbe@obspm.fr

ORCID id : https://orcid.org/0000-0002-4180-3729

**ABSTRACT**

Broad-band observations of the solar photosphere began in Meudon in 1875 under the auspices of Jules Janssen. For his part, Henri Deslandres initiated imaging spectroscopy in 1892 at Paris observatory. He invented, concurrently with George Hale in Kenwood (USA) but quite independently, the spectroheliograph designed for monochromatic imagery of the solar atmosphere. Deslandres developed two kinds of spectrographs: the "*spectrohéliographe des formes*", i.e. the narrow bandpass instrument to reveal chromospheric structures (such as filaments, prominences, plages and active regions); and the "*spectrohéliographe des vitesses*", i.e. the section spectroheliograph to record line profiles of cross sections of the Sun with a 20''-30'' spatial step. This second apparatus was intended to measure the velocities (more exactly the Dopplershifts) of dynamic features. Deslandres moved to Meudon in 1898 with his instruments and tested various combinations, in order to improve the spectral and spatial resolutions, leading to the final large quadruple spectroheliograph in 1908. CaII K systematic observations started at this date and were followed in 1909 by Hα. The service was organized by Lucien d'Azambuja and continues today. Optical and mechanical parts were revisited in 1989 and the digital technology was introduced in 2002. Full line profiles are registered for all pixels of the Sun since 2018, so that the instrument produces now data-cubes. The collection is one of the longest available (more than 100 000 observations). It contains sporadic images from 1893 to 1907 (during the development phase) and systematic observations along 10 solar cycles since 1908, in Hα and CaII K lines. This paper summarizes 130 years of observations, instrumental research and technical advances.

**INTRODUCTION**

Jules Janssen (1824-1907) founded Meudon observatory in 1875 and introduced physical astronomy in France (Malherbe, 2022). It took two decades to organize the new institution (Janssen, 1896) and build modern instruments (such as the double refractor and the one metre telescope); meanwhile, Janssen started observations of the solar photosphere, the visible layer [note 1]. The first daguerreotype of the full Sun was obtained earlier, by Hippolyte Fizeau and Léon Foucault in 1845, at Paris observatory. Janssen developed high spatial resolution photography (Lecocguen & Launay, 2005) revealing thin details of the photosphere (granules and small sunspots called pores) with an optimized refractor built by Adam Prazmowski (1821-1885). Images were recorded on 30 x 30 cm² glass plates (Figure 1). The emulsions were blue-sensitive, in the vicinity of the Fraunhofer G band (4300 Å wavelength [note 2]). 6000 plates were exposed, but 99 % were unfortunately destroyed or lost. The best observations, showing fantastic details of the granulation (the photosphere is uniformly covered by 5 million granules of 1.5'' typical size, which are the signature of the internal convection), were published in the solar photographic atlas (Janssen, 1903).

Meanwhile, Henri Deslandres (1853-1948), defended his "*Doctorat ès Sciences*" (1888) in the laboratory of Alfred Cornu at Ecole Polytechnique, and was hired in 1889 by Admiral Ernest Mouchez (the director of Paris observatory since 1878). He was in charge of organizing a spectroscopic laboratory, in the context of the development of physical astronomy initiated by Janssen at Meudon. A biography of Deslandres was published by d'Azambuja (1948) and a much more extended one by Lequeux (2022a). Deslandres first built a classical spectrograph and studied the line profiles of the ionized Calcium at 3934 Å wavelength (the K line); he succeeded to resolve the fine structure of the core in 1892. This was the starting point of the great adventure of spectroheliographs in France. The principle was outlined much earlier by Janssen (1869a, 1869b, 1869c): with a classical spectrograph, monochromatic images can be delivered by an output slit located in the spectrum (selecting the light of a spectral line) when the input slit scans the solar surface. This is possible by moving either the solar image upon the first slit, or the whole spectrograph. Janssen mentioned the spectrohelioscope made of a spectrograph rotating at high speed around the axis joining the two slits. Millochau & Stefanik (1906) described this instrument (Figure 1) as follows: *"it consists of a direct-vision spectroscope... At the focus of the telescope lens is a second slit, the two jaws of which can be independently* *adjusted and used to isolate the desired radiation. This slit is observed with a positive eyepiece. The spectroscope thus described is contained within a tube, which can be moved rapidly about its axis by means of a system of gears. This instrument thus constitutes a spectrohelioscope, and was intended for the visual study of the prominences; but by substituting a sensitive plate for the eyepiece, it might be immediately transformed into a spectroheliograph. M. Janssen's apparatus embodies the principal characteristics of the spectroheliograph, although his idea did not receive practical application during a score of years"*. Hence, two decades later, the photographic spectroheliograph was invented on this basis by George Hale in Kenwood (1892) and by Henri Deslandres in Paris (1893), simultaneously but independently. In that respect, Janssen claimed in 1906: *"the method, that I proposed in 1869, consists in the use of a second slit to select in the spectrum, formed by the first slit, a spectral line of interest. Hale and Deslandres have skilfully applied this principle"*.

Many spectroheliographs (Table 1) were built in the world, following Hale and Deslandres techniques. For instance, long series of continuous observations in the CaII K line were collected by the spectroheliographs of Kodaikanal in India, Mount Wilson in the USA, Mitaka in Japan, Sacramento Peak in the USA, Coimbra in Portugal, Meudon in France or Arcetri in Italy. They produced extended archives, some of them covering up to 10 solar cycles, which are convenient to investigate long-term solar activity and study rare events (such as energetic flares, huge sunspot groups or giant filaments [note 3]). The CaII K intensity is an excellent proxy to reconstruct past irradiance and magnetism, because the line core is greatly enhanced in bright regions (the plages) where the magnetic field and the chromospheric heating are stronger (Ermolli *et al.*, 2018).

Spectroheliographs historically produced monochromatic images along solar cycles (Figure 2) at the epoch of photographic plates. Most of them observed the CaII K and Hα Fraunhofer lines of the chromosphere. They were progressively abandoned in the second half of the twentieth century for telescopes using narrow bandpass filters (such as Fabry-Pérot or Lyot filters), which are more compact and able to observe, at higher cadence, fast evolving events (such as flares). However, with modern electronic detectors, there is a regain of interest for imaging spectroscopy, because digital spectroheliographs can now deliver full line profiles for each pixel of the Sun. This is of crucial importance for radiative transfer models and magnetohydrodynamic (MHD) simulations, which require observations of mass motions (based on Dopplershift measurements [note 4]).

**Table 1.** Major collections of CaII K line spectroheliograms. The column BW indicates the bandwidth of monochromatic images in Angström [note 2]. Meudon and Coimbra spectroheliograms have similar characteristics and select the line centre (K3) owing to their narrow waveband. For both observatories, the additional violet wing collection (K1v) doubles the indicated image number. Mount Wilson images (K3) have a similar bandpass. After Chatzistergos *et al.* (2020).

| Observatory | Detector | Period | Image number | BW[Angström] | pixel[arcsec] |
|---|---|---|---|---|---|
| Arcetri | Plate | 1931–1974 | 4871 | 0.3 | 2.5 |
| Catania | Plate | 1908–1977 | 1008 | | 1.1, 5 |
| Coimbra | Plate/CCD | 1925–2019 | 19758 | 0.16 | 2.2 |
| Kharkiv | Plate/CCD | 1952–2019 | 564 | 3.0 | 3.3 |
| Kislovodsk | Plate/CCD | 1960–2019 | 9738 | | 1.3, 2.3 |
| Kodaikanal | Plate | 1904–2007 | 45047 | 0.5 | 0.9 |
| Kyoto | Plate | 1928–1969 | 3119 | 0.74 | 2.0 |
| McMath-Hulbert | Plate | 1948–1979 | 4932 | 0.1 | 3.1 |
| Meudon | Plate/CCD | 1893–2019 | 20117 | 0.15, 0.09 | 2.2, 1.5, 1.1 |
| Mitaka | Plate | 1917–1974 | 4193 | 0.5 | 0.9, 0.7 |
| Mount Wilson | Plate | 1915–1985 | 39545 | 0.2 | 2.9 |
| Sacramento Peak | Plate | 1960–2002 | 7750 | 0.5 | 1.2 |
| Wendelstein | Plate | 1947–1977 | 422 | | 1.7, 2.6 |

This paper reviews the fantastic saga of 130 years of monochromatic observations and technological advances in Paris and Meudon observatories. Section 1 briefly presents the pioneers. Sections 2 & 3 outline the experimental and development phase (1892-1907), while Section 4 describes the second generation spectroheliographs (1908-1909). D'Azambuja's research and the synoptic maps program are described in Section 5. Specific observations (1919-1939) made for the dynamics (Dopplershift measurements [note 4]) are detailed in Section 6, while Section 7 relates the golden age of systematic observations and solar survey in the international context (1908-1988), a long period which allowed to record some of the rare events discussed in Section 8. Finally, Section 9 summarizes the modern era leading to the digital state-of-the-art instrument of 2018.

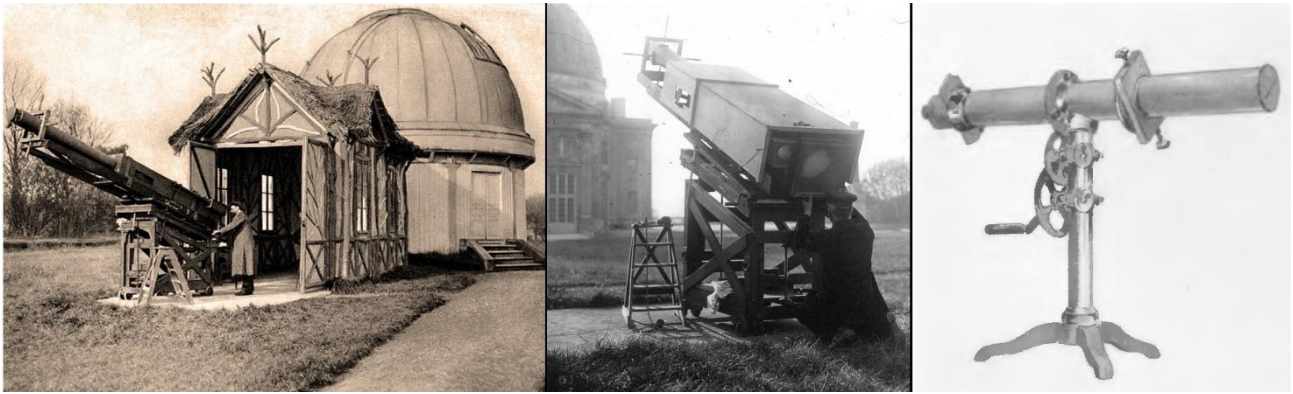

**Figure 1**. Janssen's instruments. Left and centre: the solar telescope produced 6000 images of the photosphere during 25 years. It was a broad-band refractor of 135 mm diameter (2.2 m focal length) optimized for the blue part of the spectrum. An ocular magnified the primary image (2 cm diameter) to reveal details of the solar surface on 30 x 30 cm² photographic plates. Best images of the solar granulation and sunspots were published in Janssen's photographic atlas. Right: the ancestor of the spectroheliograph is made of a direct-vision spectroscope with two slits, and rotating around its axis using a crank and gears (left: courtesy Paris observatory; centre: courtesy Gallica/BNF; right: after Millochau & Stefanik, 1906).

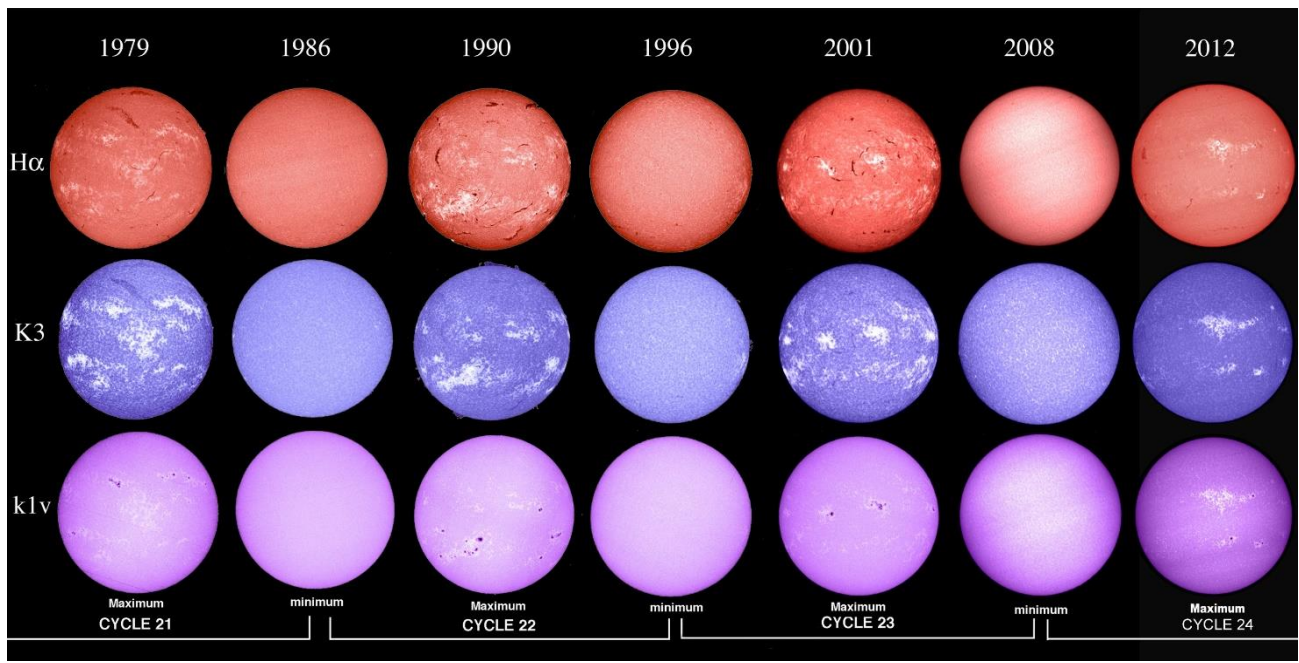

**Figure 2.** A subset of images at successive solar minima and maxima from Meudon spectroheliograph (the 11-year cycle). The chromosphere (with dark filaments and bright plages) is observed in the centre of the Hα line or in the centre (K3) of the CaII K line. The photosphere (sunspots and bright faculae) appears in the blue wing (K1v) of the CaII K line. False colours. The various wavelengths reveal solar activity phenomena at different heights in the atmosphere. The structures are related to the solar magnetism, strong in sunspots, intermediate in facular regions and weak in filaments. Courtesy Paris observatory.

# 1 THE PIONEERS

Hale founded in 1905 the Mount Wilson observatory and installed there a five-foot spectroheliograph (Hale, 1906b) fed by the Snow telescope (Hale, 1906a) to collect monochromatic images of the Sun on a regular basis. Meanwhile, Deslandres (Figure 3) moved to Meudon in 1898, which was specialized in solar physics and broad-band imagery of the Sun, in order to benefit from better conditions for the development of his research activities. Indeed, in 1908, an outstanding quadruple 14 metres spectroheliograph was built in a new laboratory. Deslandres hired in 1899 an assistant, Lucien d'Azambuja (1884-1970, Figure 2), who was only fifteen years old. He became responsible of solar instruments and organized the service of observations after the death of Janssen in 1907, when Deslandres was appointed as the new director of Meudon observatory. A biography of d'Azambuja was written by Rösch (1970) and Martres (1998). Systematic observations of the photosphere and the chromosphere [note 1] started in 1908 and continue today (Figure 2). Deslandres returned to Paris in 1926 when the two observatories gathered together, to become the first director of the enlarged institution. D'azambuja defended his "*Doctorat ès Sciences*" (d'Azambuja, 1930) and got in charge of the Meudon site. He initiated the synoptic maps of the chromosphere (d'Azambuja, 1928). He published, with his wife Marguerite (1898-1985, born Roumens, Figure 3, biography by Martres, 1991), a voluminous memoir (280 pages, d'Azambuja & d'Azambuja, 1948) about filaments and prominences [note 3], which became a reference study. This scientific work and synoptic maps were their common and major contribution to solar physics. The life at Meudon observatory, at the epoch of Deslandres was described by Lequeux (2022b), and after 1925, by M. d'Azambuja (1980). When L. d'Azambuja retired in 1954, he was succeeded at the head of the solar service by his wife, who in turn retired in 1959; they finally together left Meudon at this date, after 60 years of activity (1899-1959) devoted to the Sun.

Roger Servajean (1913-1986) and Gualtiero Olivieri (1928-2021), hired by L. d'Azambuja respectively in 1938 and 1945 (Olivieri was only seventeen years old) continued the observations, in collaboration with Marie-Josèphe Martres (1924-2017) and later Elizabeth Nesme-Ribes (1942-1996). Olivieri stayed 47 years at the observatory and had the chance to work during 14 years with the d'Azambujas. He summarized the post World War 2 (WW2) period (1945-1950) in his memoirs (Olivieri, 1993). The author of this paper is in charge of Meudon spectroheliograph since 1996, in collaboration with Isabelle Bualé and Florence Cornu.

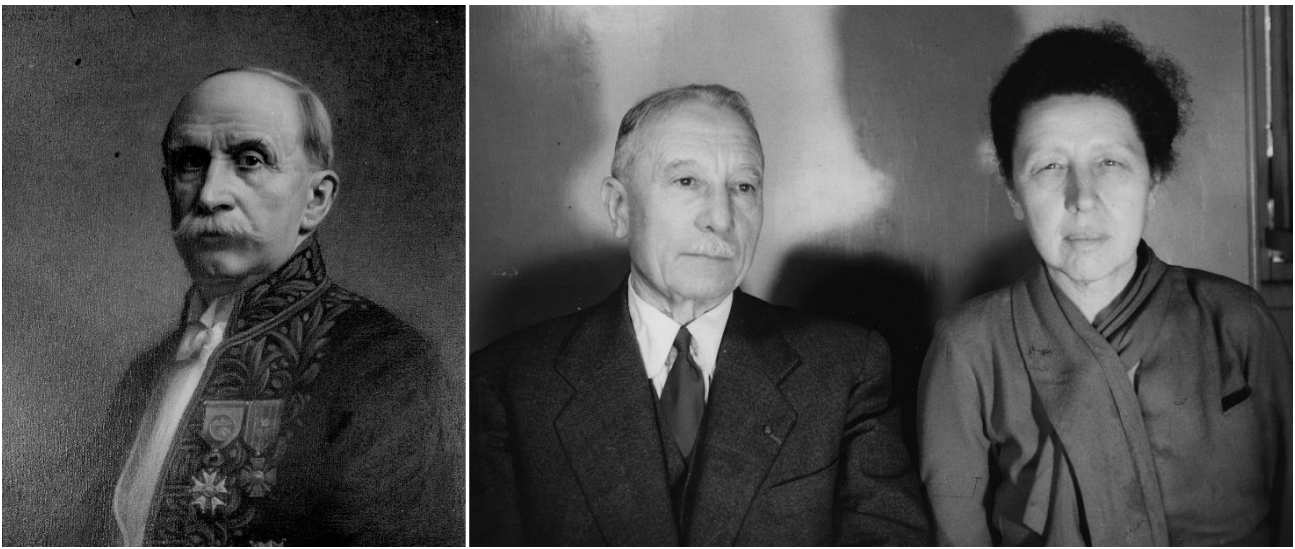

**Figure 3.** The pioneers. Henri Deslandres (1853-1948, left) developed the French spectroheliographs between 1893 and 1909. Lucien d'Azambuja (1884-1970, centre) and Marguerite d'Azambuja (born Roumens, 1898-1985, right) organized systematic observations of the Sun and created derived products (such as synoptic maps and catalogues of structures). These characters are at the origin of one of the longest world-wide collection of monochromatic images of the Sun (sporadic observations from 1893 to 1907 and continuous since 1908, apart during WW1). Courtesy Paris observatory.

## 2 THE BIRTH OF DESLANDRES SPECTROHELIOGRAPHS IN PARIS (1891-1897)

Deslandres was hired in 1889 to establish a spectroscopic laboratory by Admiral Mouchez, the successor of Urbain Le Verrier (1811-1877) at the head of Paris observatory. Le Verrier was a well known specialist of celestial mechanics, but he neglected the importance of physical astronomy under development in many institutions and countries. Dollfus (2003a, 2003b) related the beginnings of solar spectroscopy in Paris. Deslandres (1894a, 1894b) described in details his instrumental research. He first mounted a classical spectrograph in the northern garden of Paris observatory (Figure 4) fed by a small Foucault siderostat of 30 cm diameter [note 5]. Many tests of different optical combinations were undertaken. In May 1891, he used a 30 cm diameter objective to form the solar image on the slit of his spectrograph. It was composed of a 0.50 m collimator and 0.70 m chamber with an adjustable grating, a photographic plate carrier, and a second auxiliary chamber for visual inspection of the spectrum. In August 1891, other tests were made with a 20 cm diameter concave mirror, first with prisms, and later with a more dispersive Rowland plane grating in order 4. At last, in December 1891, Deslandres introduced a longer focal length collimator and chamber (1.35 m for both), allowing to double the dispersion. As early as 1891, he was able to observe the Hydrogen Balmer series (Hα to Hε, or electronic transitions from level 2 to levels 3, 4, 5, 6, 7) as well as the strong CaII H and K lines in the near ultraviolet part of the spectrum. As photographic emulsions were blue sensitive, he studied the CaII K line profiles of various chromospheric features [note 3] such as bright plages, prominences and sunspots (Figure 5). Deslandres noticed the complexity of the Calcium line profiles, composed of a narrow core (K3), two adjacent peaks (K2r, K2v) and extended wings (K1r, K1v, letters r and v referring respectively to the red and violet wing), which are formed at different altitudes (Figure 5, top). He saw that the K2 and K3 components may vary considerably, depending on the observed structure, in particular in plages (the line is very sensitive to temperature fluctuations).

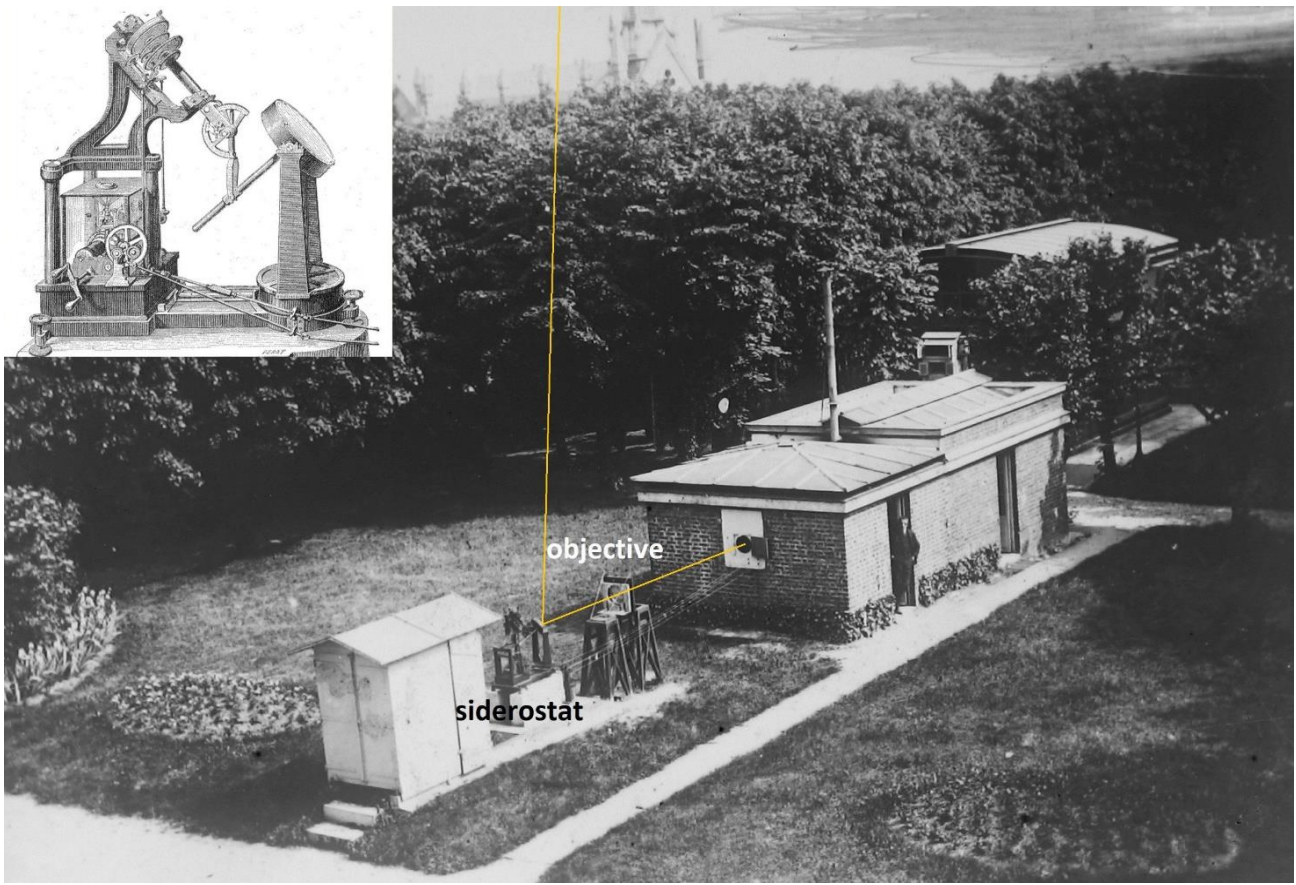

**Figure 4.** Deslandres's spectroheliograph in Paris (1890-1897), in the garden located north of the observatory. The solar light is collected by a small Foucault siderostat (upper corner). The reflected beam is horizontal and has a fixed direction. It crosses an objective which forms the solar image on the spectrograph slit, located inside the house. The spectrograph and the photographic plate translate together on rails to scan the solar surface and record monochromatic pictures. Courtesy Paris observatory.

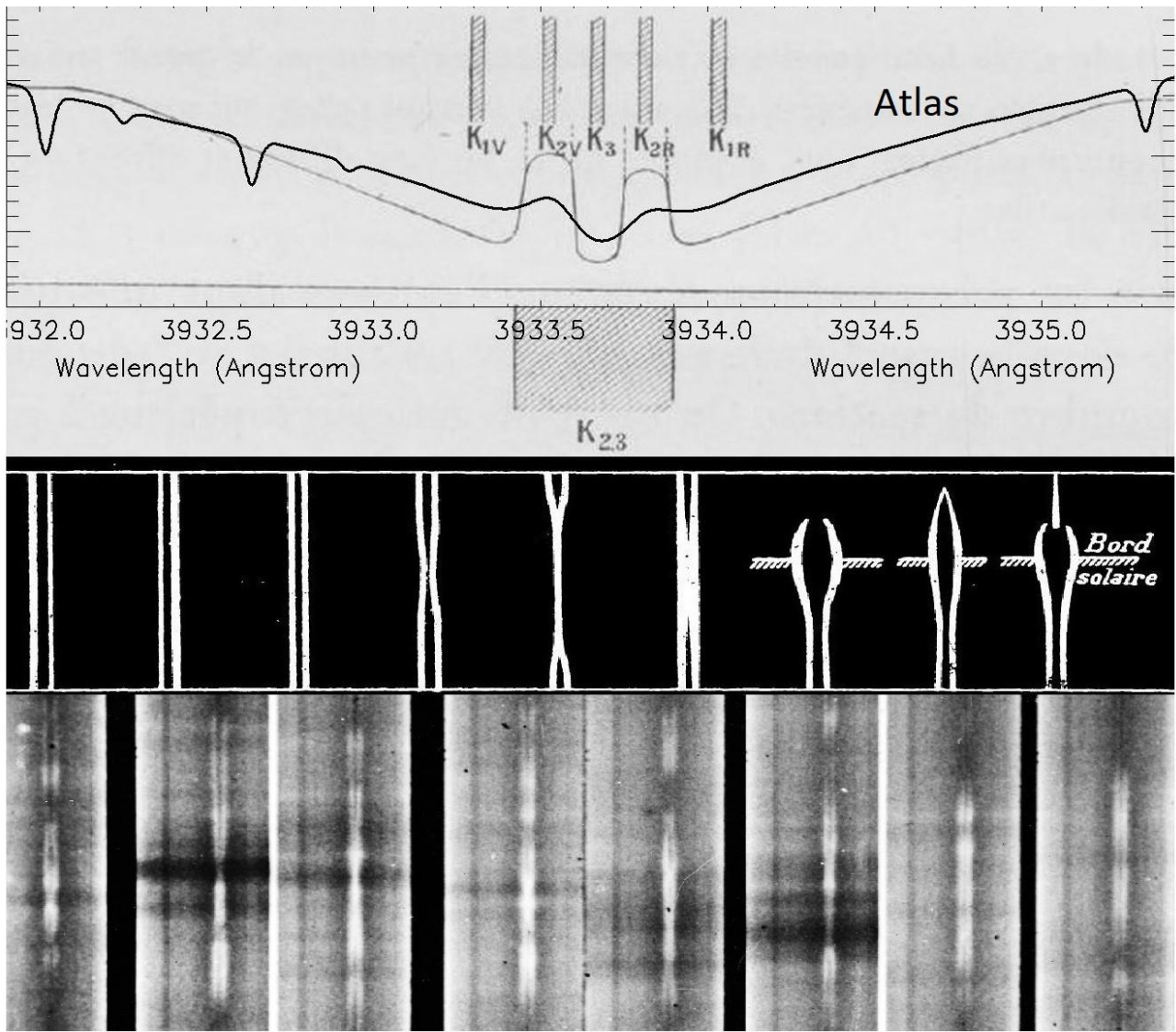

K2v, K2r drawings and K line observations at various locations

**Figure 5.** CaII K line observations by Deslandres (1892) and identification of the line components (grating spectrograph). Top: drawing showing the details of the K line profile (a modern atlas profile is superimposed, after Delbouille *et al.*, 1973). K1v, K1r are respectively the violet and red wings, formed in the photosphere. K2v and K2r are peaks at ± 0.2 Å from the line centre (K3), all are formed in the chromosphere, but K3 has the highest altitude. The first spectroheliograph used a single prism and was unable to separate K2v, K3 and K2r; it provided instead a wavelength integration called K23 over the 0.5 Å bandwidth. Middle: drawings of the line profiles at various places of the Sun, including the limb (white lines represent the K2 components). Bottom: spectra recorded on photographic plates (18 May 1894) in various active regions (wavelength in abscissa, slit direction in ordinates). The K2 peaks are particularly bright in (hot) plages and vanish in (cold) dark sunspots. After Deslandres (1905a and 1910) and courtesy Paris observatory.

The first spectroheliographs were tested between 1892 and 1894 with the CaII K line, because of the blue sensitivity of photographic plates. The initial goal of Deslandres was to investigate the nature of solar features, and he decided to determine two different parameters: the intensity and the velocity of the structures. For that purpose, he constructed and tested two kinds of spectroheliographs (Deslandres, 1905a).

The first instrument, called "*spectrohéliographe des formes*", produced a monochromatic image with the second slit in the spectrum selecting the K23 central part of the K line (i.e. the wavelength integration of K2r, K3, K2v, see Figure 5), showing the shape (the "*formes*" in French) and the brightness of chromospheric structures; this instrument was moving continuously on rails to scan the Sun with the first slit. The entrance objective had a diameter of 12 cm, 2.80 m focal length, while those of the collimator and chamber were respectively 0.50 m and 1.0 m (magnification 2.0), so that the monochromatic image had a diameter of 50 mm. The dispersive element was a single prism of light flint and 60° angle at minimum deviation.

The second instrument, called "*spectrohéliographe des vitesses*" or section spectroheliograph, used a much larger slit in the spectrum (typically 1.5 mm instead of 0.1 mm), in order to transmit the full line profile of CaII K (Figure 5, bottom). It allowed to measure the Dopplershift [note 4] of the K3 component in order to quantify the line-of-sight (or radial) velocities of the structures. As the output slit was large, the spectrograph had to move step by step (1 step = 20''-30'' on the solar surface), so that line profiles corresponding to equidistant cross sections of the Sun were registered contiguously on the photographic plate. In order to resolve properly the details of line profiles, a high dispersive element was used (Rowland grating at order 4). The entrance objective had a diameter of 30 cm, 5.0 m focal length, while those of the collimator and chamber were identical (1.30 m, magnification 1.0), so that the height of the cross sections was about 50 mm.

D'Azambuja (1920a, 1920b) explained the mechanisms involved in these first spectroheliographs (Figure 6). The spectrograph moves at constant speed $V_1$ to scan the solar surface (first slit $F_1$). The second slit ($F_2$) selects the spectral line. In order to form a monochromatic image of the Sun, the photographic plate ($C_2$) translates simultaneously at speed $V_2 = \gamma V_1$, where $\gamma = f_2/f_1$ is the magnification factor of the spectrograph. This is the role of the two levers ($L_1$) and ($L_2$), connected to the fixed point (A) and rotating around the pivot point (R) attached to the spectrograph. The lengths ($L_1$, $L_2$) of the levers are in the ratio of the optical magnification, so that $L_2/L_1 = \gamma$. The levers (L1, L2) and points (A, R) are indicated in Figure 6 on the two photographs showing the "*spectrohéliographe des formes*" (left, with continuous motion and low dispersion prism) and the more sophisticated "*spectrohéliographe des vitesses*" or section spectroheliograph (right, with discontinuous translation and high dispersion grating). Step motions of the second instrument were done manually, or by the mechanical device described in the next section.

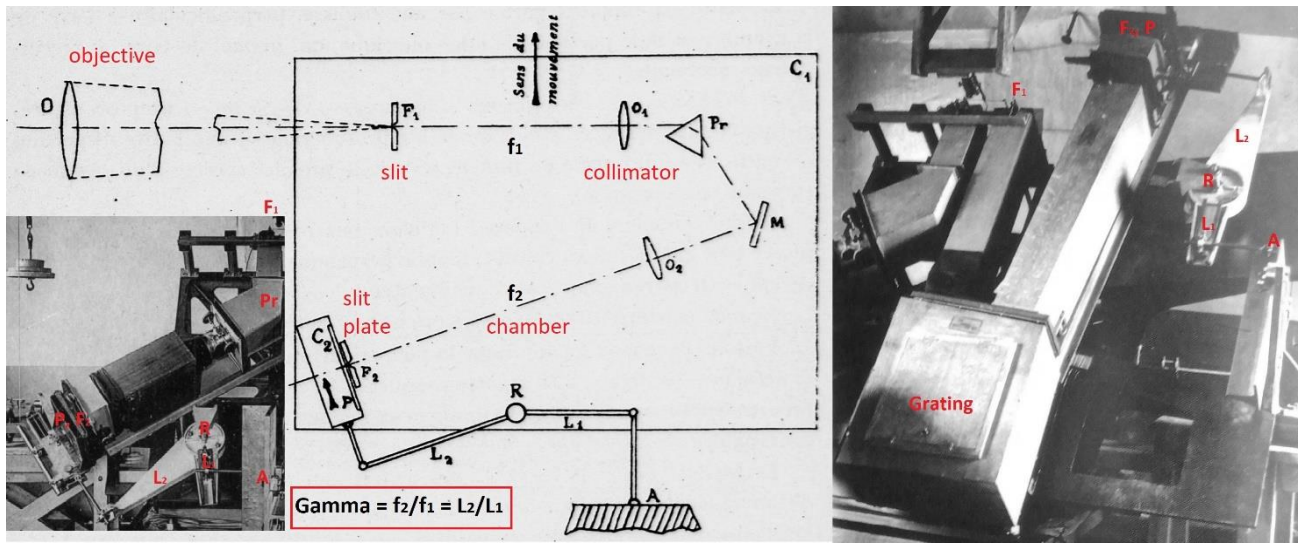

**Figure 6.** The principle of spectroheliographs (drawing after d'Azambuja, 1920b). The spectrograph moves on rails and the slit $F_1$ scans the solar surface. The second slit ($F_2$) selects the line in the spectrum. A mechanism translates the photographic plate to form a monochromatic image (O = objective; $F_1$ = entrance slit; $O_1$ = collimator; $O_2$ = chamber; M = flat mirror; Pr = prism; $F_2$ = output slit; A = fixed point; R = pivot point attached to the spectrograph). The optical magnification $\gamma = f_2/f_1$ is equal to the ratio $L_2/L_1$ of the lever lengths. The photographs show the first spectroheliograhs (left: low dispersion prism instrument; right: high dispersion grating instrument) installed on the same rolling table; in both cases, the points (A, R) and the levers ($L_1$, $L_2$) are indicated. Courtesy Paris observatory.

Some of the first spectroheliograms (1893) obtained in the K23 centre of the CaII K line, using a single prism, are displayed in Figure 7 (image diameter 50 mm). They revealed sunspots and bright areas around, called by Deslandres "*plages faculaires*". Hale used the name of "*flocculi*", also for small bright points forming the chromospheric network outside active regions. Deslandres preferred the terms "*plages*" or "*faculae*" for large bright areas, which depend on the phase of the 11-year solar activity cycle, and reserved "*flocculi*" for the quiet network, which is permanent. Dark solar filaments [note 3] were not yet visible, because the K23 waveband was too broad. We know that these structures only appear in the narrow K3 line core, which could not be selected by this first generation spectroheliographs. However, K23 revealed prominences above the limb. As the required exposure time for prominences is five times the one used for the solar disk, Deslandres put a circular mask upon the image to remove the excess of light entering the spectrograph.

In 1897, this spectroheliograph was a bit modified. Hereafter, it was fed by the same 30 cm diameter entrance objective (5.0 m focal length) than the one used for the section spectroheliograph (described below). The collimator and chamber lenses, of respectively 0.65 m and 1.0 m focal length (optical magnification 1.54), produced larger monochromatic images (82 mm diameter) which will become later the standard size.

The first section spectroheliograms were obtained in 1894 with the second instrument moving discontinuously by steps of 30'' and using a high dispersion Rowland grating to resolve the CaII K line profile (Figure 8). The image was composed of 60 steps, or 60 cross sections of the Sun, providing the line profiles along the entrance slit. For each step of the spectrograph (0.8 mm), the photographic plate was displaced of 2.2 mm. For that reason, the final final image (about 50 x 140 mm²) was elliptic and composed of the juxtaposition of 60 spectra (λ, y), λ and y being respectively the wavelength and the coordinate along the slit. The width of the output slit in the spectrum (1.5 mm) corresponded to a spectral band of about 2-3 Å centred on the line. The displacements of the line core (K3 Dopplershifts [note 4]) could be measured to provide the line-of-sight velocity of material.

We saw that Deslandres's initial aim was to produce monochromatic images of the chromosphere and also to investigate mass motions in solar structures. This first generation instruments developed in Paris (1891-1897) demonstrated that this goal could be achieved. Many developments continued in Meudon after 1898, and led in 1908 to the second generation of instruments with improved capabilities, for both systematic and research observations.

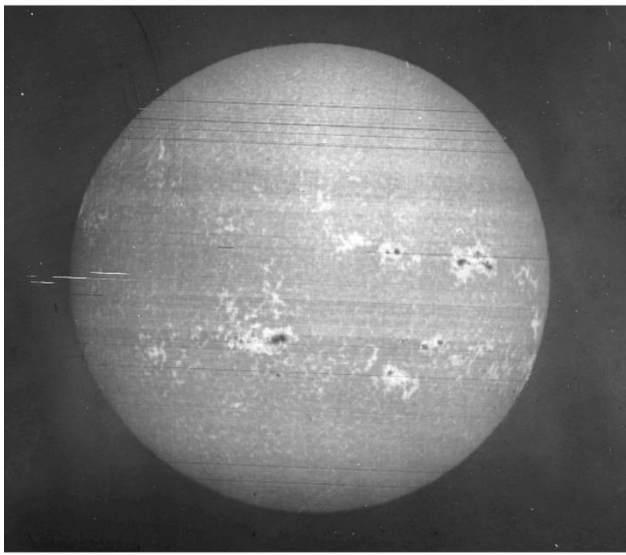
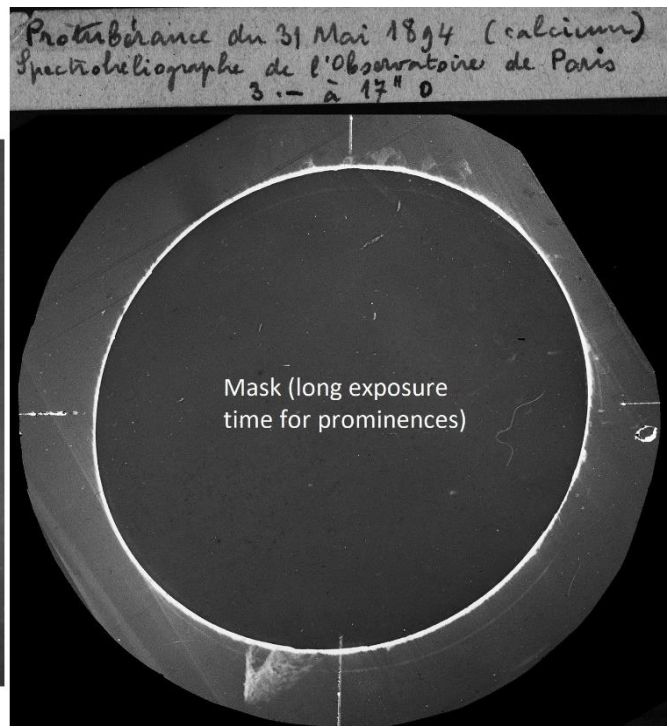

**Figure 7.** First spectroheliograms (1893) obtained in Paris by Deslandres in the CaII K line (K23 or wavelength integration of K2v, K3, K2r components). In the right picture, a mask has been put in the solar image to increase the exposure time and reveal solar prominences at the limb, which are much fainter than the solar disk. Courtesy Paris Observatory.

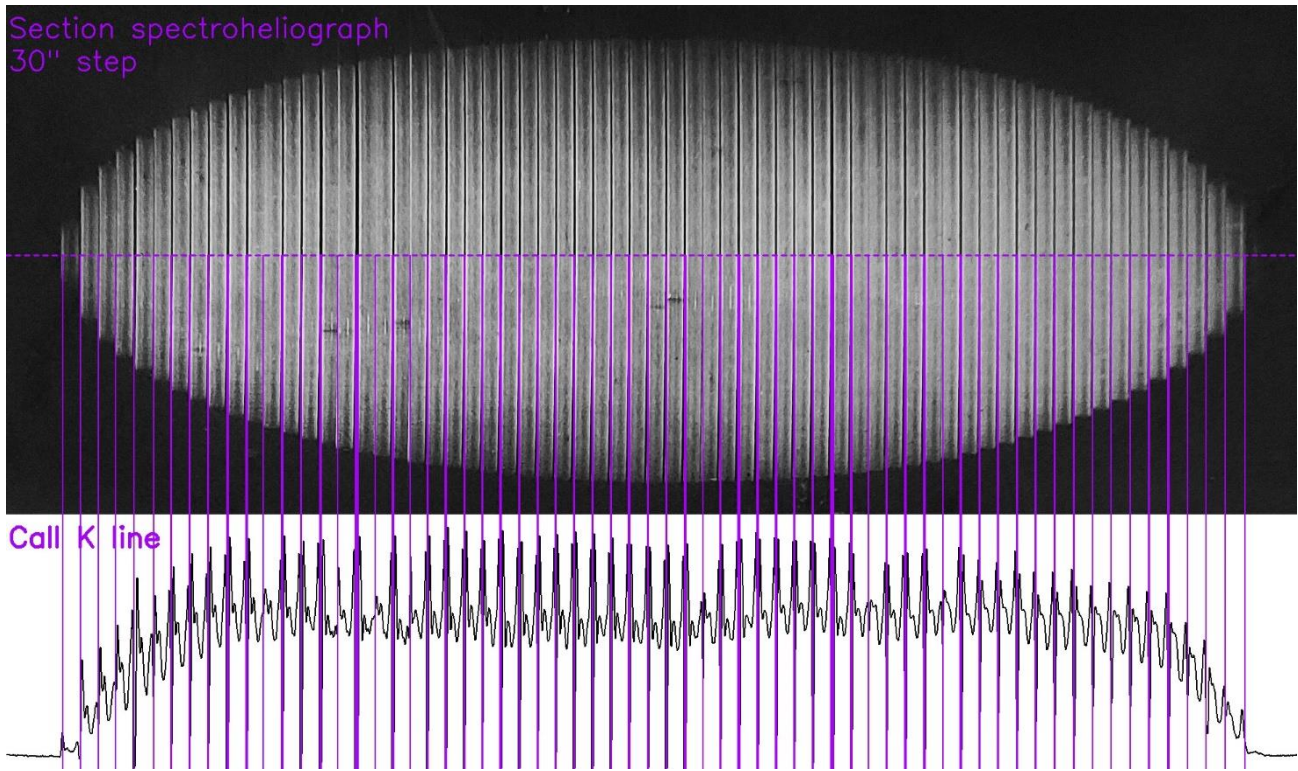

**Figure 8.** Top: first section spectroheliogram (18 May 1894) obtained in Paris by Deslandres in the CaII K line. The scan of the solar surface was discontinuous; the slit moved by steps of about 30'', so that 60 cross sections of the Sun were recorded on the photographic plate. The output slit, located in the spectrum, was enlarged to register the line profile over a 2-3 Å waveband allowing the measurement of line-of-sight velocities, via the Dopplershift of the K3 component. Bottom: the CaII K line profiles along the equator for each section. Courtesy Paris observatory.

**3 DESLANDRES EXPERIMENTS AT THE MEUDON "PETIT SIDEROSTAT" (1898-1907)**

Deslandres decided to quit Paris in 1898 and relocate his instruments, and research activities, at Meudon observatory. Indeed, he wrote: "*it was not possible to do more in Paris observatory, which is mainly involved in astrometry and considers solar researches as a second priority… I moved in 1898 to Meudon observatory which is devoted to solar physics and offers more interesting perspectives…*". But at Meudon, it was not so simple and everything had to be done ! When Deslandres arrived, the observatory was as shown by Figure 9 and Figure 10. The small laboratory (B) was built in 1898 in order to install Paris instruments, and the much larger laboratory (D) was erected in 1907 for the second generation of spectroheliographs.

At Deslandres's arrival in Meudon, Paris instruments were installed in the new laboratory of Figure 11 called "*Petit Sidérostat*" (small siderostat in French), which was built for that purpose (Deslandres, 1905b). The solar light was caught by a polar siderostat (Figure 12) [note 5] previously used by Janssen for the Venus transit of 1874 in Japan. The diameter of the mirror was 30 cm. In practice, Deslandres found in Meudon worse observing conditions than in Paris. The polar siderostat brought many difficulties in comparison to the Foucault siderostat available in Paris; in particular the spectroheliographs could not be horizontal but had to be inclined. However, the context (as solar physics was the priority) and the perspectives for future extensions and developments were much better.

Deslandres was soon helped by L. D'Azambuja, hired in 1899. He wrote: "*the new organization was made available in Meudon with a young assistant, L. d'Azambuja, whose zeal and intelligence were remarkable*". The two spectroheliographs (Figure 13) were mounted on an inclined table of 1.70 x 1.10 m², containing the polar axis, and moving on two rails with rollers. The 20 cm diameter imaging objective had a focal length of 3.10 m.

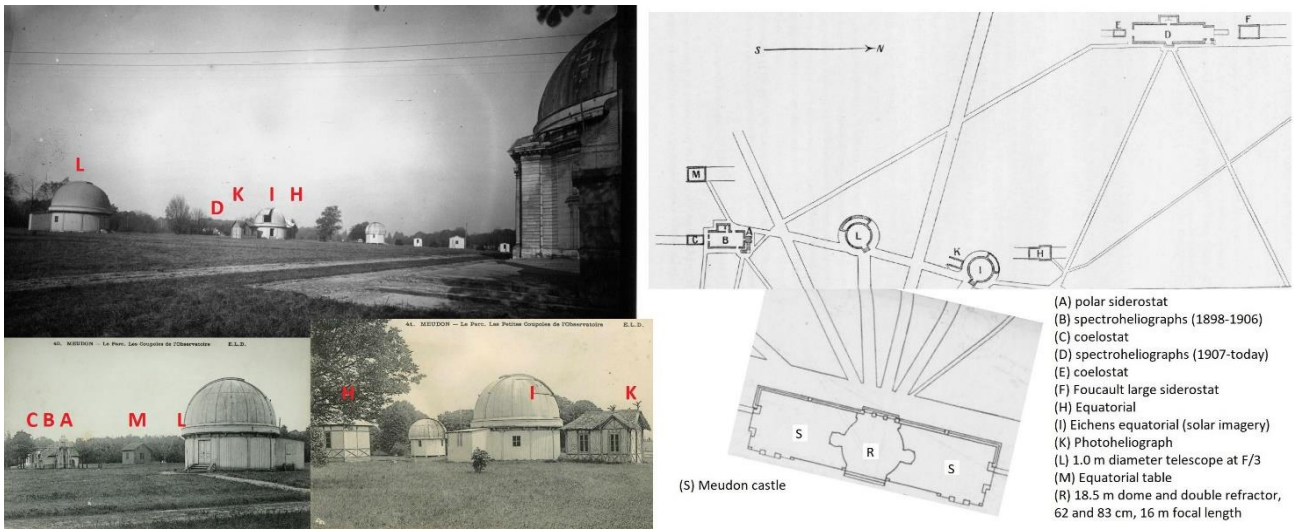

**Figure 9.** Meudon observatory in 1910. Deslandres installed Paris instruments in the (B) laboratory, built in 1898. It was fed by a polar siderostat (A); in 1906, a small coelostat (C) was added. In 1908, the large quadruple spectroheliograph was installed in the building (D), fed by the coelostat (E). Later a large Foucault siderostat (F) was added but it was never used for daily observations. The famous Janssen's refractor for broadband solar imagery is in (K). The night instruments (83 cm and 62 cm double refractor, 100 cm telescope), completed in 1893, are respectively in (R) and (L). Courtesy Paris observatory and after Deslandres (1910).

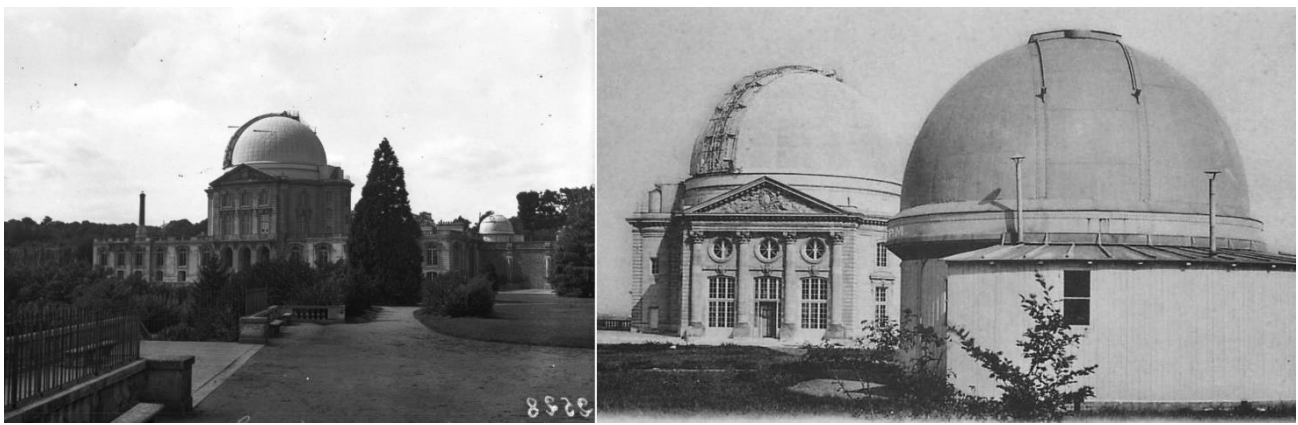

**Figure 10.** Meudon chateau (left: Est side ; right: West side) and the night instruments, indicated respectively by letters R (double refractor) and L (100 cm telescope) in Figure 9. Courtesy Gallica/BNF (left) and Paris observatory (right).

The classical spectroheliograph for imaging the calcium chromosphere (Figure 14, left) moved continuously during the observations, under the action of the hydraulic motor (the piston in cylinder a, b and water reservoirs a', b' of Figure 13). The spectrograph had focal lengths of 0.30 m and 1.0 m, respectively for the collimator and the chamber lenses (magnification 3.3), and used a single 60° flint prism (moderate dispersion of 0.06 mm/Å). The thin slit in the spectrum delivered K1v (violet wing) or K23 monochromatic images of 90 mm diameter, respectively for the photosphere and the chromosphere. Exposure times were in the range 2-6 minutes, and three times longer for prominences. Filaments, which require to select only the CaII K3 line core, were of course not yet visible.

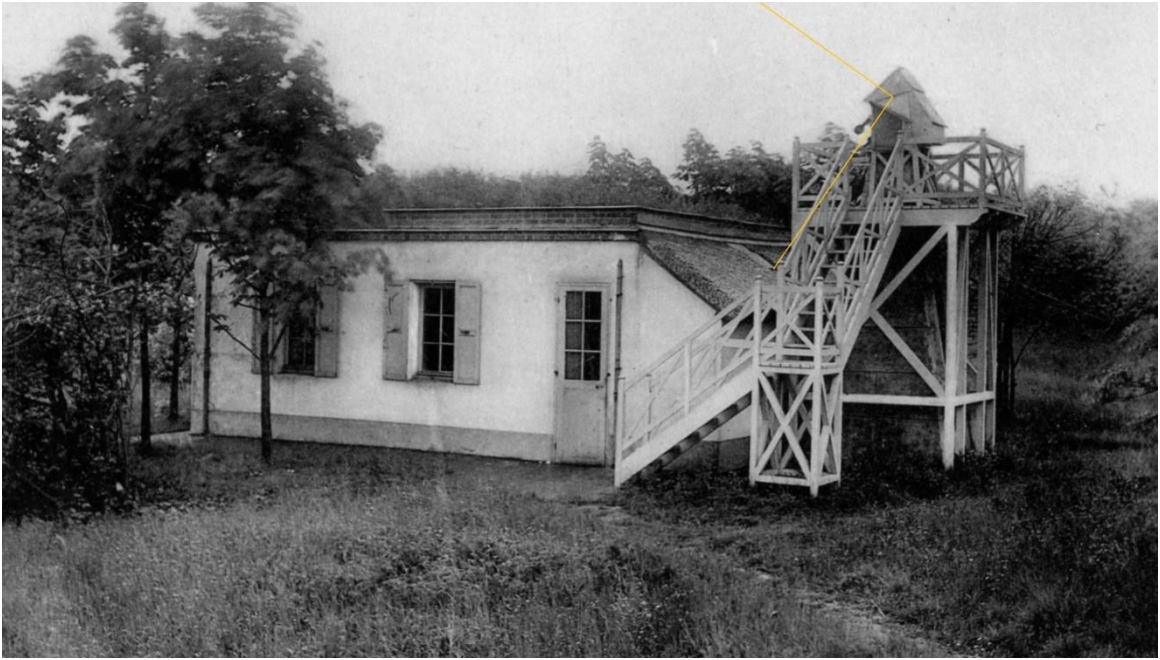

**Figure 11.** When Deslandres moved to Meudon in 1898, this laboratory, called "*Petit Sidérostat*" (B in Figure 9), was erected. The solar light was caught by a polar siderostat (A in Figure 9) and directed towards an imaging objective (see details of Figure 12), located near the rotating flat mirror. A hole in the roof allowed the converging light beam to feed the two spectroheliographs installed on an inclined table inside the house. Courtesy Paris observatory.

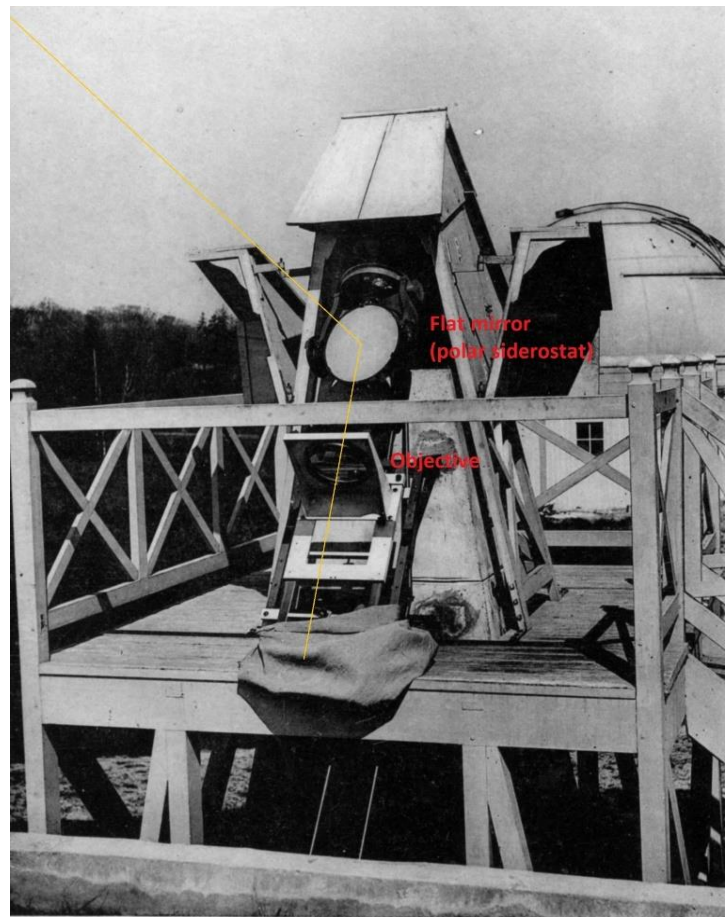

**Figure 12.** The polar siderostat (A in Figure 9) and the imaging objective (in the background, one sees the dome of the 100 cm telescope). The converging light beam was directed towards the spectroheliographs through a hole in the roof of the house. It delivered a primary 30 mm solar image, enlarged by the spectrographs. Courtesy Paris observatory.

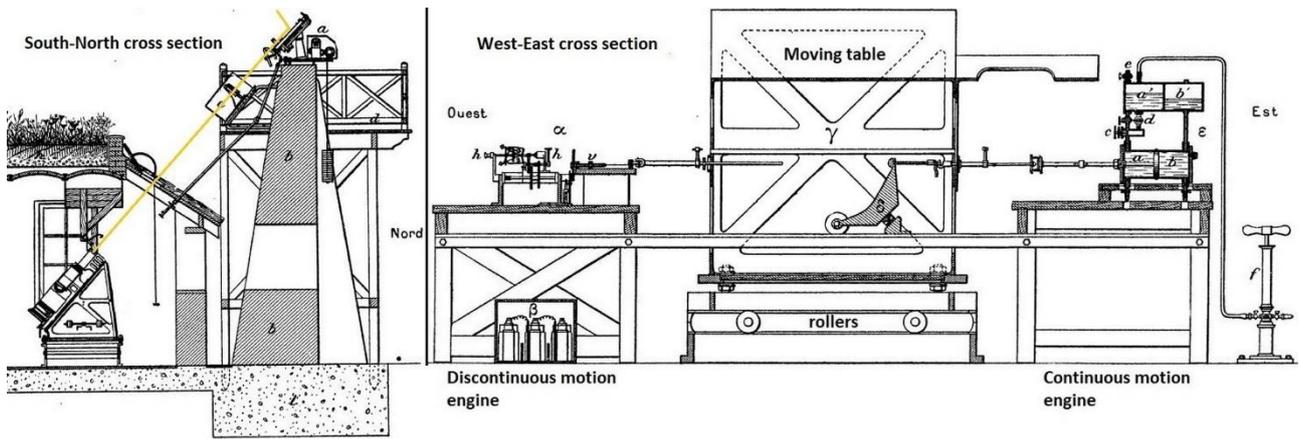

**Figure 13.** Side and face views of the spectroheliographs installed by Deslandres in the "*Petit Sidérostat*" laboratory (1898). Left (side view): the polar siderostat (a), the imaging objective (c) and the spectroheliographs. Right (face view): the spectroheliographs are fixed upon an inclined table (γ), moving on rollers, and containing the polar axis. The first engine (East) is for the continuous translation needed for CaII K spectroheliograms (the "*spectrohéliographe des formes*"), which moved under the action of the hydraulic motor (ε) and the piston between water reservoirs (a, b). The second mechanism (West) is for the step motion required for the section spectroheliograph (the "*spectrohéliographe des vitesses*") dedicated to radial velocities, using the electric motor (α), the step engine (h) of Figure 15 and the set of batteries β. After Deslandres, 1905 b.

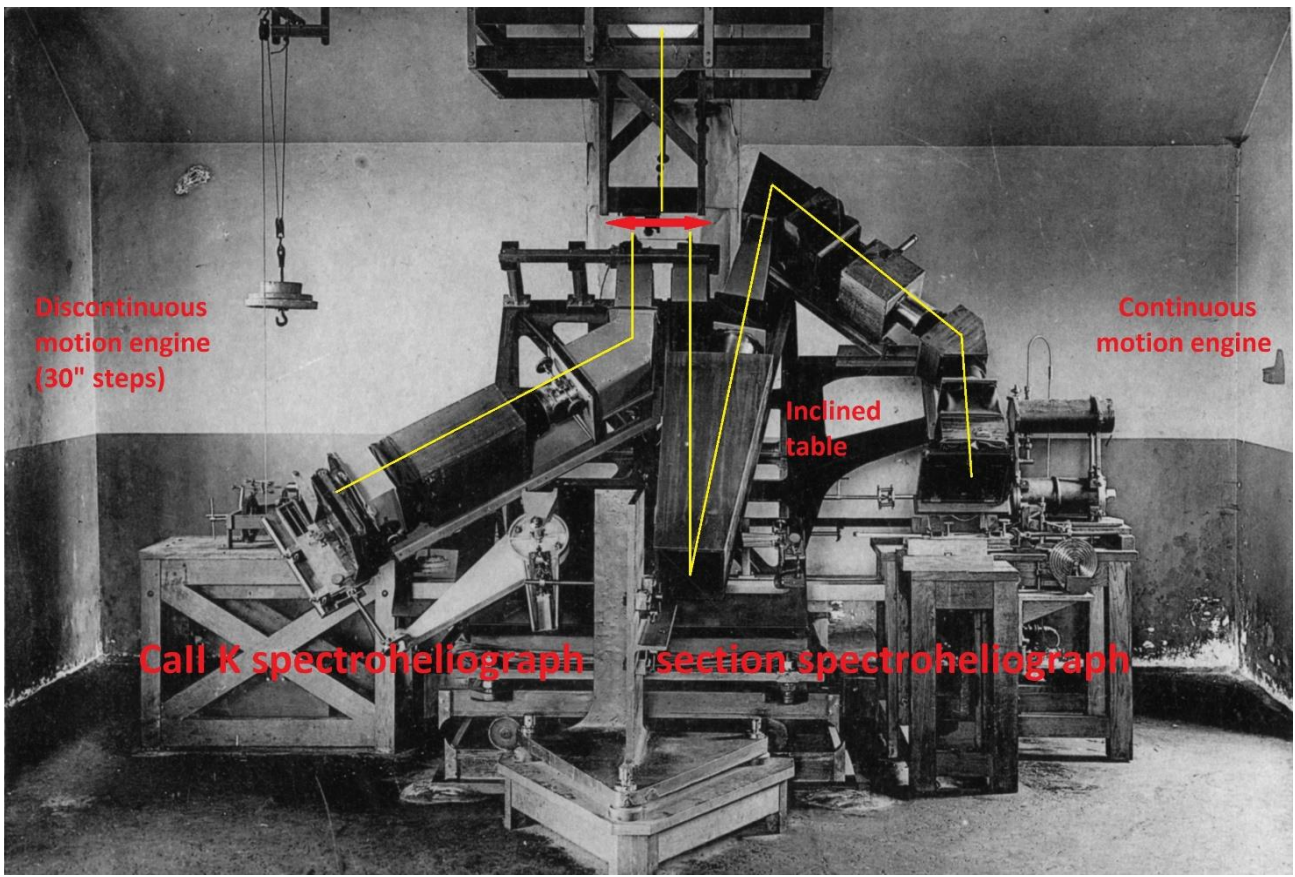

**Figure 14.** The two spectroheliographs of the "*Petit Sidérostat*" house in Meudon. Left: the low dispersion CaII K (K23) spectroheliograph used a single prism and moved at constant speed during observations. Right: the section spectroheliograph moved discontinuously by steps of 30'', and had a higher dispersion. It was made of two additive spectrographs (details in Figure 15), the first one with a plane Rowland grating and the second one with a prism to separate orders and attenuate scattered light. Courtesy Paris observatory.

The section spectrograph (Figure 14, right, and Figure 15, left) was improved in comparison to the one operating in Paris. It used a 55 x 80 mm² Rowland grating at order 4 (560 grooves/mm, five times more dispersive than the prism), a 0.50 m collimator and a 1.0 m chamber (optical magnification equal to 2.0). It recorded 90 contiguous solar cross sections of 2 x 60 mm² on the photographic plate. About 10 seconds of exposure time were necessary for each section and 15 minutes for the full observation. However, the grating created a lot of scattered light and mixed orders, so that a second and low dispersive spectrograph, with a single 60° prism and a third slit, was added in order to eliminate parasitic light and record good quality spectra. This second spectrograph had a magnification of 1.0 (0.50 m focal length, both for the collimator and the chamber). Moving the spectroheliograph table automatically step by step was a complicated task. Deslandres imagined the special engine of Figure 15 (right) which produced step motions from gears rotating at constant speed under the action of an electric motor. The translation of the table took 1.5 s per step, while the exposure time was selectable in the range 1.5-16.5 s for each section using various gear combinations.

The two spectroheliographs used the system of levers described above (Figure 6), in order to synchronize the translation of the photographic plate carriers with the spectrographs, but of course observations with both instruments were always performed successively.

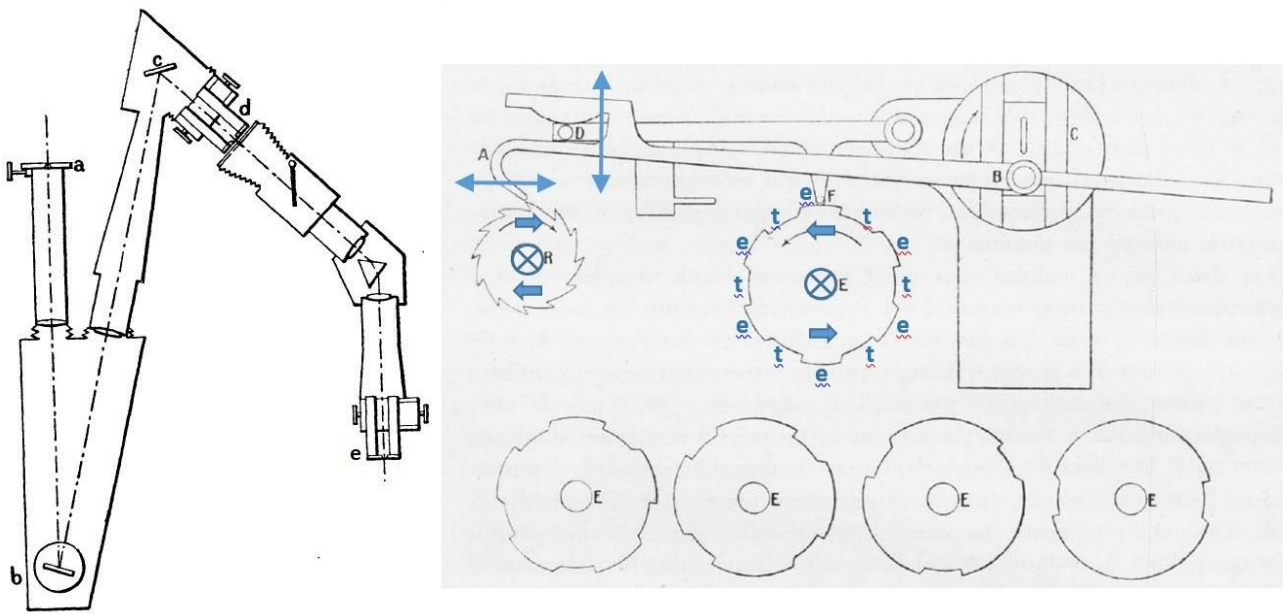

**Figure 15.** The section spectroheliograph ("*spectrohéliographe des vitesses*") of the "*Petit Sidérostat*" in Meudon. Left: entrance slit in the image plane (a), Rowland grating (b), plane mirror (c), second slit (d) in the spectrum (a few Å bandpass); the second spectrograph with the prism eliminates parasitic orders and reduces the scattered light; the photographic plate is located at the third slit (e). Right: the step engine converts the continuous rotation of wheels (C) and (E) into a step rotation of gear (R) under the combined motions of elements (AB) and (DF), respectively in horizontal and vertical directions. The spectroheliograph was pushed by a screw mounted on the (R) axis. When index (F) was high (e position), the exposure of the plate occurred; when (F) was low (t position), the spectrograph moved under the action of the hook (A). Several wheels (E) corresponded to various exposure times: 6, 5, 4, 3 or 2 teeth provided respectively 1.5, 2.1, 3.0, 4.5 or 7.5 s exposure times. Each step took 1.5 s, and the rotating period of wheel (E) was always 18 s. After Deslandres (1905b, 1910).

The location of the two spectroheliographs in the laboratory are reported in Figure 16 and denoted (a, b). However, the section spectroheliograph (b) for radial velocities often changed, and Deslandres tested many optical combinations, with different chambers, number of slits (2 or 3), gratings and series of prisms (Dollfus, 2005). In particular, the two spectroheliographs could not work simultaneously to record the structures (requiring the continuous motion of the table) and spectra (step by step), and a research instrument to study many other lines than CaII K was lacking to work independently. For that purpose, a small coelostat (Figure 16) [note 5] was added in 1906 south of the building, which was enlarged to receive the new instruments (c, d) which prefigured the second generation spectroheliographs of 1909: contrarily to (a, b) which moved to scan the solar surface, spectrographs (c, d) were standing still, horizontal and fed by a moving objective. The coelostat directed the light towards a concave mirror (m) of 20 cm diameter and 4.0 m focal length. It formed a solar image of 37 mm on the entrance slit of either the spectroheliograph (c) or (d). The scan of the solar surface was ensured by the translation of the imaging concave mirror (m). Instrument (d) used a grating and also concave mirrors, both for the collimator (1.0 m) and the chamber (2.50 m), providing spectra of 92 mm height. It was a multi-purpose spectroheliograph for lines in a wide range of wavelengths from the yellow to the violet.

Instrument (c) is detailed in Figure 17; it was a section spectroheliograph, but the imaging mirror (m) was replaced by a 18 cm diameter lens (3.60 m focal length), providing spectra of 140 mm height with the 0.80 m collimator and 3.40 m chamber. The final image was composed of 140 sections of 1 mm, so that it did not deliver elliptic images (as the one of Figure 8), but circular images, contrarily to the initial instrument (b). This spectroheliograph could work at the same time than the instrument (a) recording the solar structures and shapes. Hα spectra were obtained in 1908 with the Rowland grating, and the series of three prisms (62°30' angle) was mainly used to measure the Dopplershifts of the CaII K3 component, from which line-of-sight velocities were deduced.

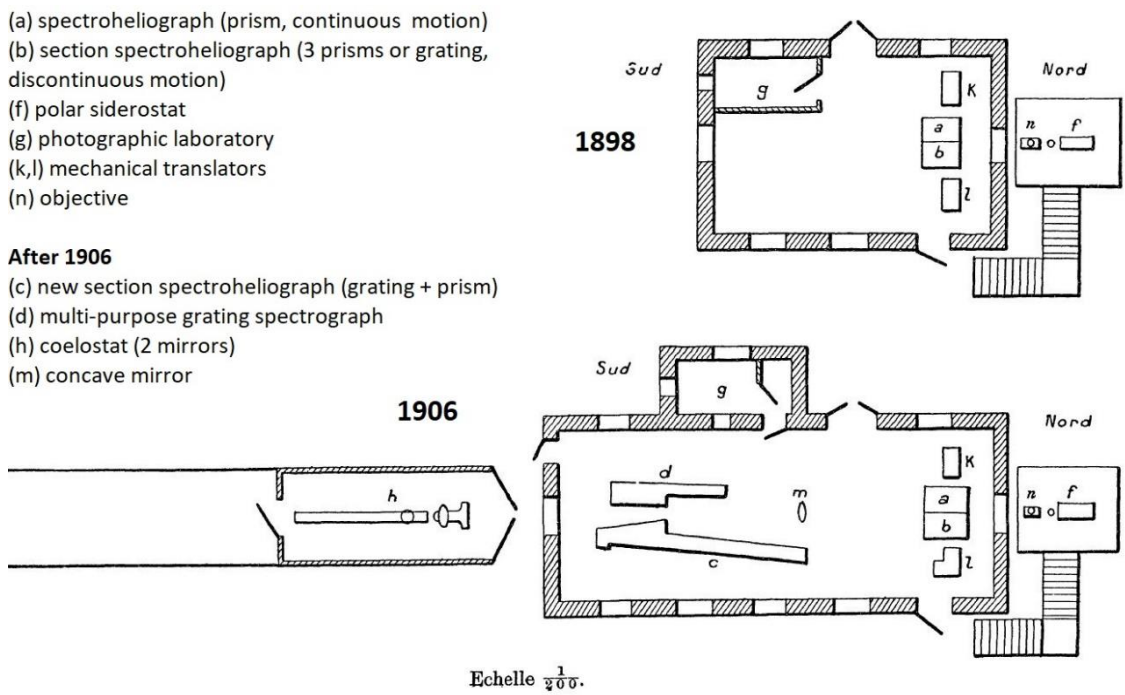

**Figure 16.** The "*Petit Sidérostat*" in 1898 (top, 5 x 7 m²) and in 1906 (bottom, 5 x 12.5 m²). The 1898 organization corresponds to the transfer of the classical spectroheliograph (a) and section spectroheliograph (b) from Paris to Meudon, using the polar siderostat (f) and objective (n). The 1906 organization included the two experiments (c, d) and the laboratory was enlarged. (d) was a multi-purpose and multi-wavelength research spectrograph, while (c) was an improved section spectroheliograph mainly dedicated to CaII K velocity measurements. Both were fed by the new coelostat (h), south of the house. After Deslandres (1907a).

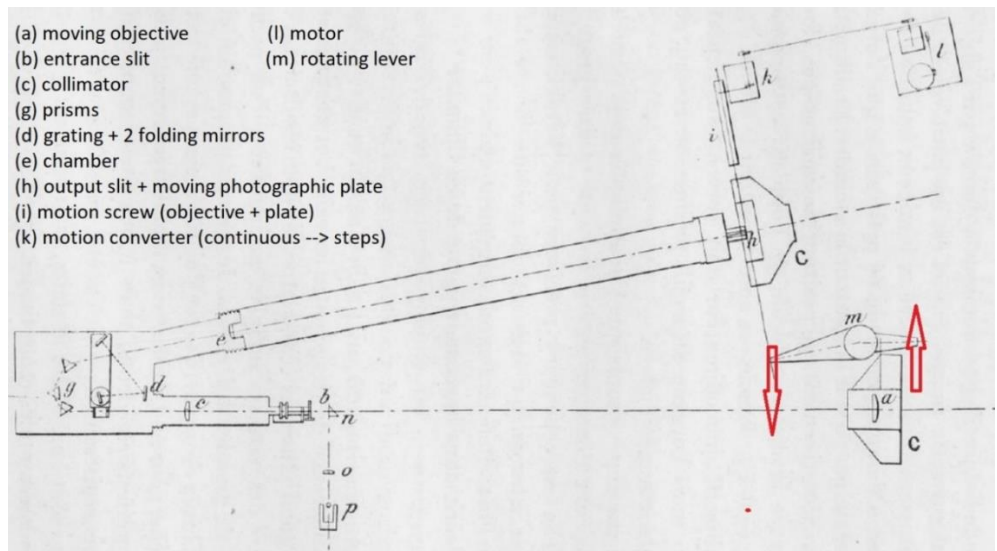

**Figure 17.** The new section spectroheliograph at the "*Petit Sidérostat*" in 1906, denoted (c) in Figure 16. It uses either three prisms for CaII K (g) or a plane grating for Hα. (l) is a motor acting simultaneously on the entrance objective (a) and the photographic plate (h) through the (k) mechanism, which converts the continuous rotation of the motor (l) into step motions (details in Figure 15). (i) is a rotating screw translating the plate; (m) is a fixed pivot point with two levers of different lengths, in order to adapt the ratio of the translation velocities of (h) and (a) to the optical magnification of the spectrograph (as in Figure 6). After Deslandres (1910).

## 4  THE QUADRUPLE SPECTROHELIOGRAPH (1908-1909) OF DESLANDRES AND D'AZAMBUJA

The first congress of the International Union for the Cooperation in Solar Research occurred in Meudon chateau in May 1907 (Figure 18), a few months before the death of J. Janssen. A preparatory meeting took place three years before at Saint Louis, in the USA, and was organized by G. Hale (who founded previously the Mount Wilson observatory, where the Snow telescope and spectroheliograph were installed). Two other meetings were held at Mount Wilson (1910) and Bonn (1913). This was the starting point of a large international cooperation in solar astronomy, which developed mainly after WW1. The International Research Council (IRC, renamed in 1931 ICSU (International Council of Scientific Unions) promoted the cooperation through a series of international unions, such as the International Astronomical Union (IAU) and the International Union for Geodesy and Geophysics (IUGG), both created in 1919. The first meetings took place in Rome (1922) and Cambridge (1925). A special commission between both unions was formed to study the Sun-Earth relations. Deslandres was member of the executive committee which recommended to coordinate solar and terrestrial observations. He wrote in 1925 that the quadruple spectroheliograph "*can be presented as the typical instrument for solar studies*". Let us now describe this sophisticated achievement.

At the death of Janssen (1907), Deslandres was appointed to the position of director of Meudon observatory. He got a financial support to build a 6 x 26 m² laboratory (called the "*Grand Sidérostat*", Figure 19) for his project of large quadruple spectroheliograph (Figure 20 and Table 2). A large Foucault siderostat (75 cm diameter) located north of the housing was ordered to catch and follow the Sun, and reflect the light in the horizontal direction. But assembling a so big and sophisticated mechanical piece was a difficult undertaking which took much more time than expected. For that reason, Deslandres decided to install in 1908, south of the laboratory, a two mirror coelostat (40 cm diameter) to feed the new generation instruments. It was much easier and faster to set up, and this temporary coelostat finally became permanent. The hand-made diagrams of the coelostat and housing can be found in Deslandres's notebook (Deslandres, 1907b); the coelostat was replaced in 1920 by a new one. The quadruple spectroheliograph (Figure 20) was described by Deslandres (1910) and d'Azambuja (1930). Deslandres wrote: "*It is a complex instrument, difficult to adjust. In my research, I have been permanently helped by L. d'Azambuja, a young and brilliant astronomer, whose name is associated to mine in my discoveries*". Indeed, Deslandres was more and more busy with the tasks of Meudon director, and delegated the installation and adjustments of the new quadruple spectroheliograph to d'Azambuja. The goal was to have two instruments for systematic observations in Hα and CaII K, and two multi-purpose instruments with improved capabilities (such as higher dispersion and larger spectral domain, from the ultraviolet to the near infrared), dedicated to solar research in spectroscopy.

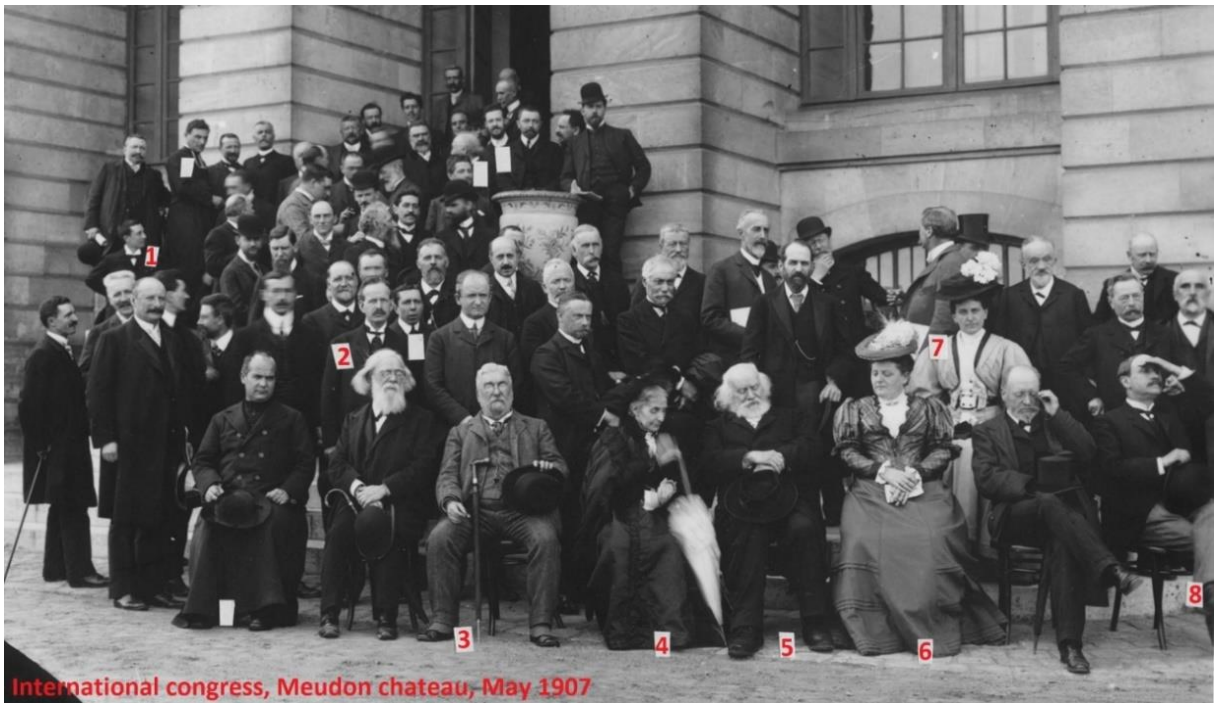

**Figure 18.** The first congress of the International Union for the Cooperation in Solar Research in May 1907 at Meudon chateau. Some people are identified on the photography: (1) L. d'Azambuja, (2) G. Hale (3), B. Baillaud (director of Paris observatory), (4) Mrs Janssen, (5) J. Janssen (director of Meudon observatory), (6) Janssen's daughter, (7) Mrs Deslandres, (8) H. Deslandres. The characters of the photography were identified by d'Azambuja (1967) in a letter sent to R. Michard. Courtesy Paris observatory.

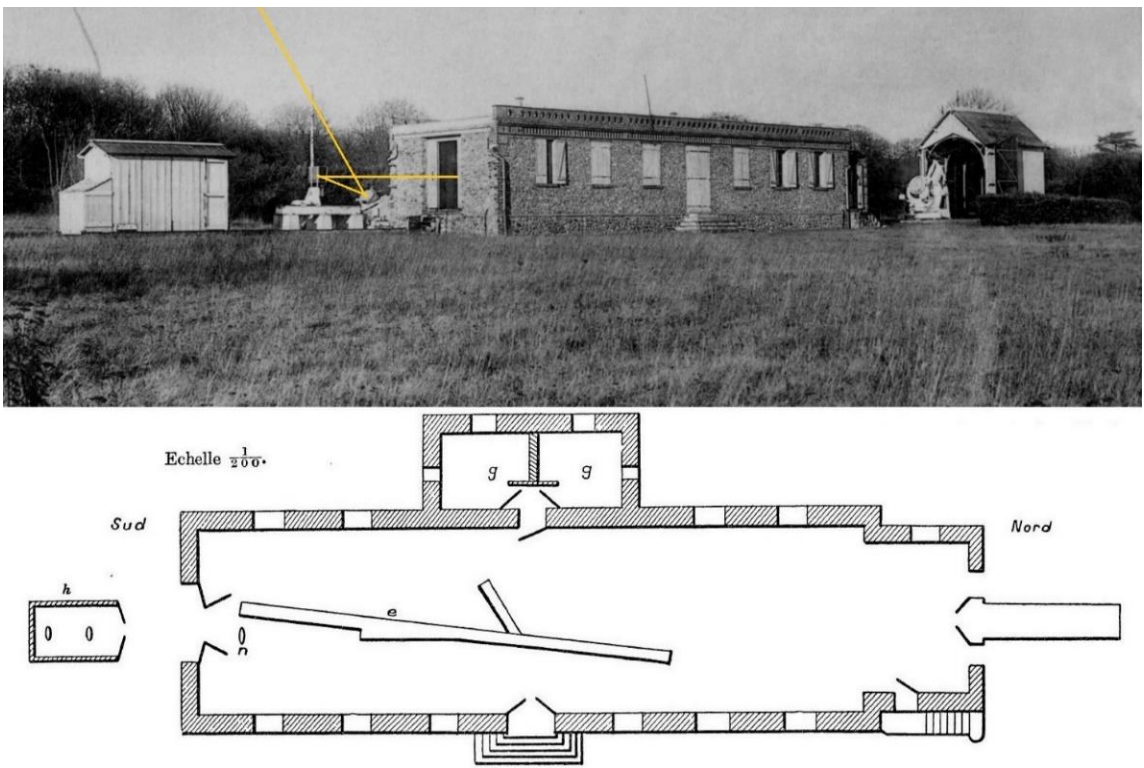

**Figure 19.** The "*Grand Sidérostat*" laboratory built in 1907 (6 x 24 m²) allowed to start systematic observations of CaII K in 1908 and in Hα in 1909, with two dedicated combinations of the large quadruple spectroheliograph (e). It was fed by the coelostat (h) south of the building (left) and objective (n). The Foucault siderostat (right) was completed later, so that it was never used for systematic observations. (g) were photographic cabinets. After Deslandres (1907a) and courtesy Paris observatory.

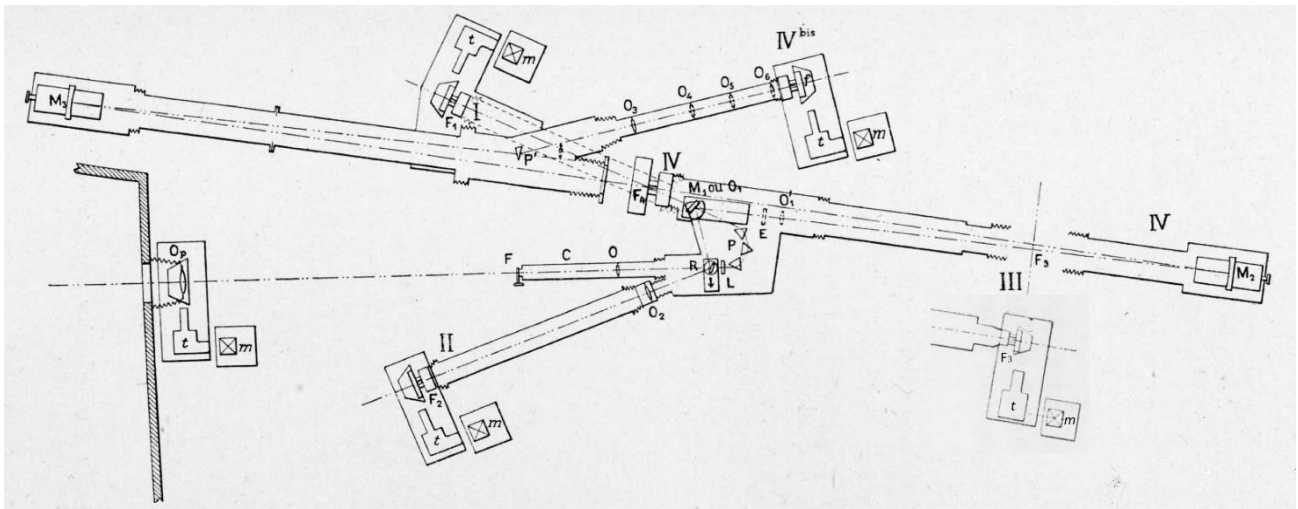

**Figure 20.** The quadruple spectroheliograph (1909) was extremely complex, versatile and composed of four instruments more or less superimposed. Two of them (spectrographs 1-2) were used for systematic observations, alternatively in CaII K and Hα, and the others (spectrographs 3-4-4bis), more dispersive, for research purpose. The details of each instrument are reported in Table 2, and their optical design in Figure 22 and Figure 26. The entrance objective (Op), slit (F) and collimator (O) were common to the four combinations, so that they could not work simultaneously, but alternatively. However, it was easy to switch between them. After d'Azambuja (1930).

**Table 2.** The optical combinations of the quadruple spectroheliograph of Figure 20 (optical designs detailed in Figure 22 and Figure 26). The entrance objective, slit and collimator were common to the four chambers.

| Chamber | Optical Path (Figure 20) | Length (m) | Image diameter (mm) | Spectral line |
|---|---|---|---|---|
| **For systematic observations** | | | | |
| 1 | $P+O_1+F_1$ | 3.0 | 86 | CaII K |
| 2 | $R+O_2+F_2$ | 3.0 | 86 | Hα |
| **For research observations** | | | | |
| 3 | $(P\ or\ R)+M_1+O'_1+F_3$ | 3.0 | 86 | multi-purpose |
| 4 | $(P\ or\ R)+M_1+M_2+F_4$ | 7.0 | 205 | multi-purpose |
| 4bis (2 spectros and image reduction γ) | $(P\ or\ R)+M_1+M_2+F_4+M_3+P'$ | 2 x 7.0 | | |
| | with $O_1$ (γ = 0.29) | | 58.5 | multi-purpose |
| | or $O_2$ (γ = 0.19) | | 39.5 | multi-purpose |
| | or $O_3$ (γ = 0.09) | | 19.5 | multi-purpose |
| | or $O_4$ (γ = 0.04) | | 9.0 | multi-purpose |

A picture of the quadruple spectroheliograph, built by Deslandres and d'Azambuja, is shown in Figure 21. On the foreground are the entrance objective and the two 3.0 m spectroheliographs (1-2) dedicated to systematic observations, which were organized by d'Azambuja and started respectively in 1908 for CaII K and 1909 for Hα. On the background are located the spectroheliographs devoted to solar research (3-4-4bis), respectively of 3.0, 7.0 m and 2 x 7.0 m length. All combinations used alternatively the 25 cm imaging objective (4.0 m focal length), scanning the solar surface and providing an image of 37 mm, the entrance slit (S) and the collimator (13 cm diameter, 1.30 m focal length).

The spectroheliographs (1-2 of Table 2) dedicated to systematic observations are displayed in Figure 22. A plane grating was used for Hα (568 grooves/mm, size of 8 x 5.5 cm²) while a series of three prisms (flint glass at minimum deviation, 61° angle, section of 15 x 13.5 cm²) was used for CaII K. The entrance slit had a width of 35 µm (corresponding to 1.8'' or 1300 km on the Sun). The second slit (80 µm width) in the dispersed spectrum selected narrow wavebands of 0.14 Å and 0.47 Å, respectively for CaII K and Hα lines. With the 2.31 magnification factor, the chamber lenses (15 cm diameter, 3.0 m focal length) provided standard 86 mm monochromatic images on the photographic glass plates.

Contrarily to the first generation instruments of Paris and Meudon, the quadruple spectroheliograph was motionless and there was no longer any mechanical coupling (such as the system of Figure 17) between the entrance objective and the four photographic plates. Each carrier (the [t, m] couples of Figure 20) had its own synchronous electric motor connected to an adjustable "*transformateur des vitesses*", some kind of velocity converter (Figure 23), in order to guarantee circular monochromatic images. The roller of this mechanical device allowed to adapt carefully the plate speed to the scanning speed of the Sun, the ratio of which depending on the various magnifications (0.04-5.54) of the spectrographs. Indeed, as indicated by Table 2, the spectroheliograms had diameters in the range 9-205 mm (86 mm for standard observations), for a common entrance image of 37 mm delivered by the imaging objective.

Figure 24 shows some of the first systematic observations, which started in 1908 for the CaII K line (the central K3 component was selected, as it was now possible to isolate it with the series of three prisms) and in 1909 for Hα. However, during a short period, the old and new generation of spectroheliographs, in the two different laboratories, worked together, as shown by Figure 25. CaII K1v and K23 were produced with the low dispersion, single prism spectroheliograph of the "*Petit Sidérostat*", while K2v and K3 were observed with the high dispersion, three-prisms instrument of the "*Grand Sidérostat*". In particular, dark solar filaments appeared for the first time on K3 spectroheliograms, owing to the narrow waveband of 0.14 Å (Figure 25). About 1000 observations were performed per year. For instance, in 1920, the number of observations was 159 K1v, 104 K23, 285 K3, 168 long exposures for prominences and 112 Hα (828 total). After WW1, there was a lack of manpower and observations with the old instruments were progressively abandoned; only K1v (for sunspots and faculae) and K3 (for active regions, bright plages and dark filaments) continued with the new generation instruments, together with Hα which reveals another aspect of filaments and plages.

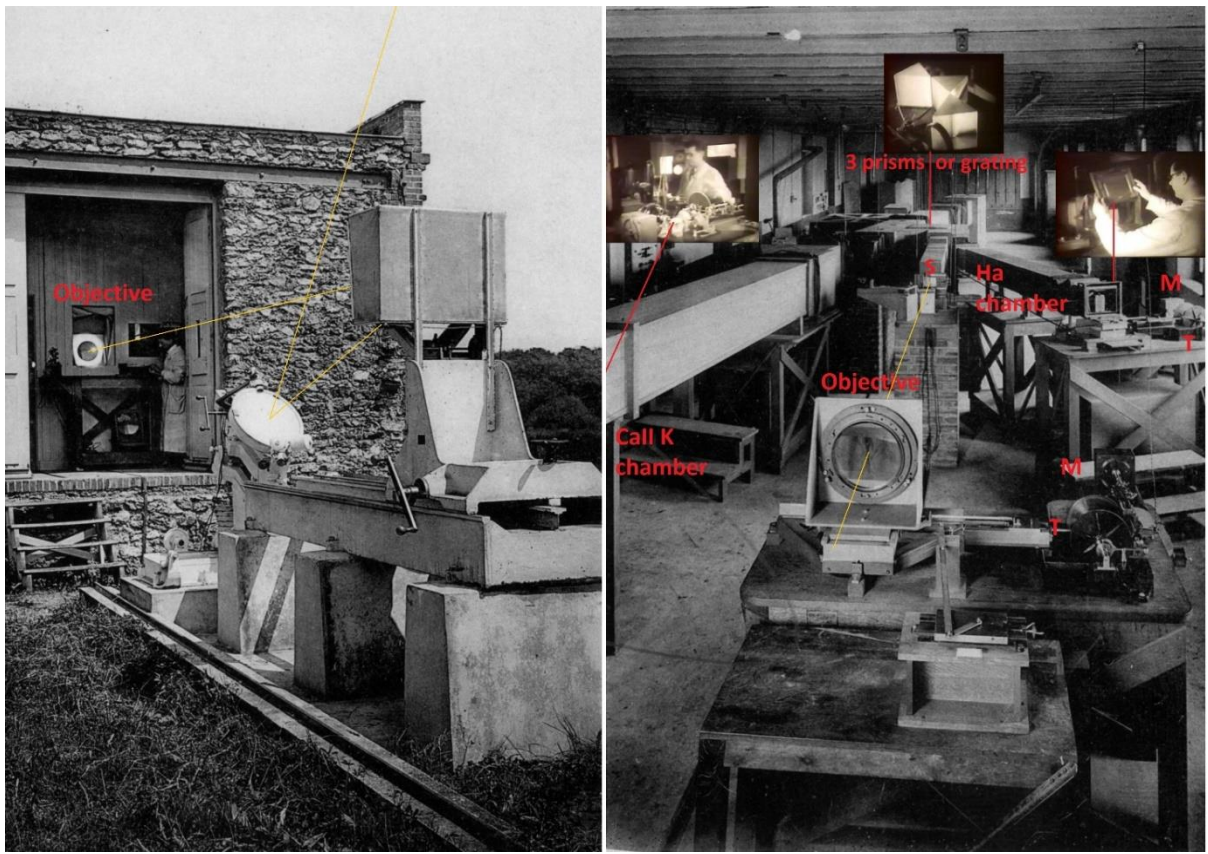

**Figure 21.** The quadruple spectroheliograph (1909) with the coelostat (left) and the imaging objective (4 m focal length). One sees the CaII K chamber (left arm, design of Figure 22), the Hα chamber (right arm, design of Figure 22), the electric motors (M) moving the objective and the photographic plates, and the associated velocity transformers (T, details in Figure 23). (S) is the common entrance slit. Courtesy Paris observatory.

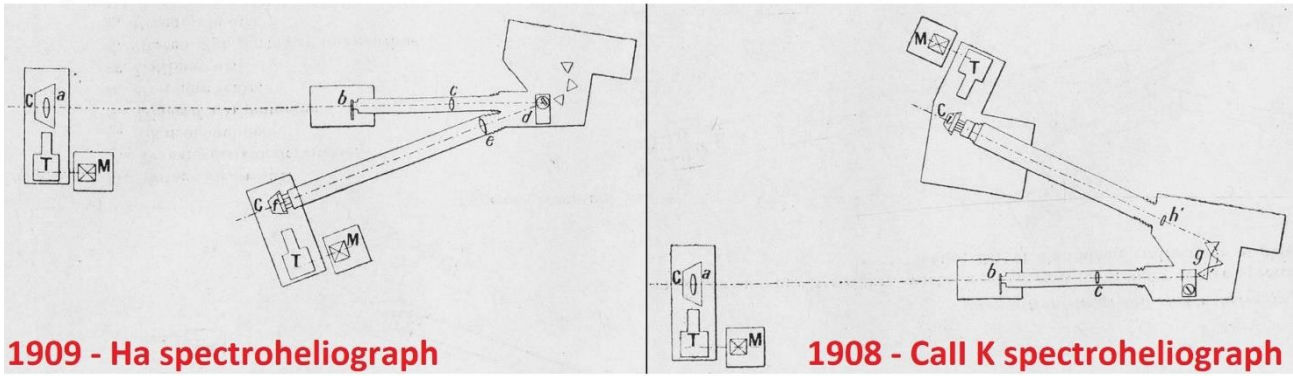

**Figure 22.** The two spectroheliographs (1-2) dedicated to systematic observations in Hα (left, with plane grating) and CaII K (right, with three prisms). (a), (b), (c) are common elements (respectively the imaging objective, slit and collimator). Each spectral line has its own chamber. M and T are couples of synchronous motors and velocity transformers (details in Figure 23), which adjust precisely the speeds of the imaging objective and the photographic plates. After Deslandres, 1910.

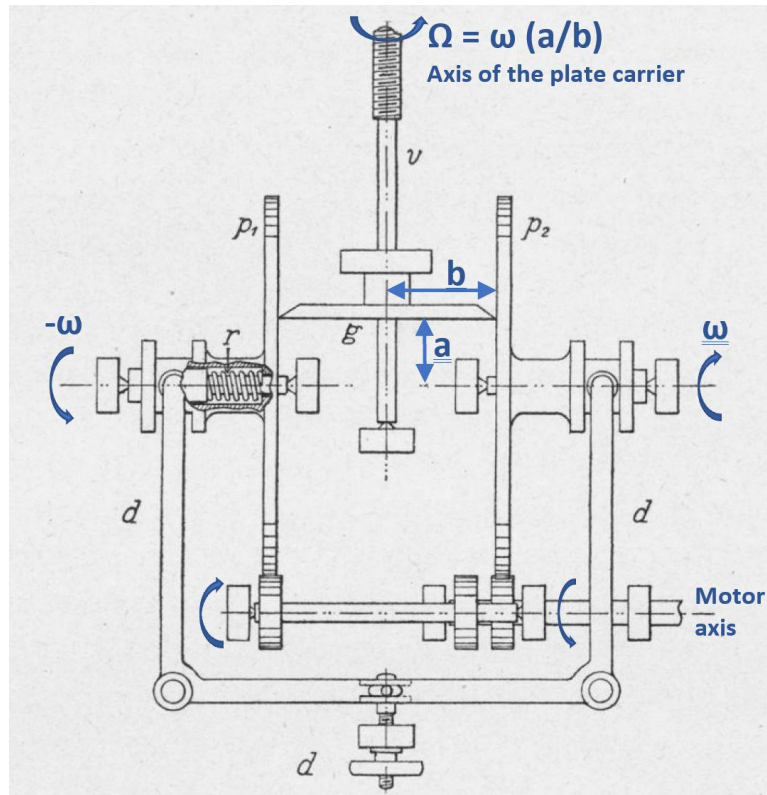

**Figure 23.** The mechanical "*transformateur des vitesses*" (velocity transformer or converter) is the major element to synchronize the translation of the imaging objective and the photographic plates. (v) is the screw (adjustable angular velocity Ω) pushing or pulling the plate carrier. (p1) and (p2) are circular disks rotating in opposite directions (fixed angular velocities ±ω) under the action of the motor. (g) is a roller rotating at speed Ω = ω (a/b) where (b) is fixed and (a) is the adjustable distance between the roller and the rotation axis of (p1) and (p2). The value of (a) is chosen with the precision of 0.1 mm in order to get a circular monochromatic image on the photographic plate; the length (a) is changed by relaxing the spring (r) with the lever (d). After Deslandres, 1910.

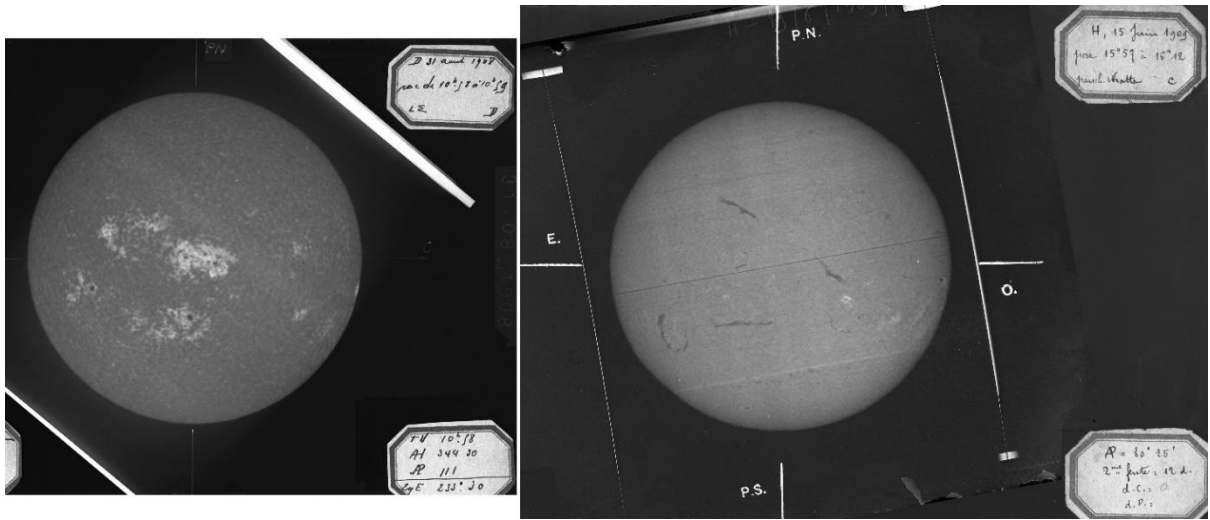

**Figure 24.** Some of the first systematic observations made with the two spectroheliographs of Figure 22. Left: CaII K3 (31 August 1908). Right: Hα (15 June 1909). Courtesy Paris observatory.

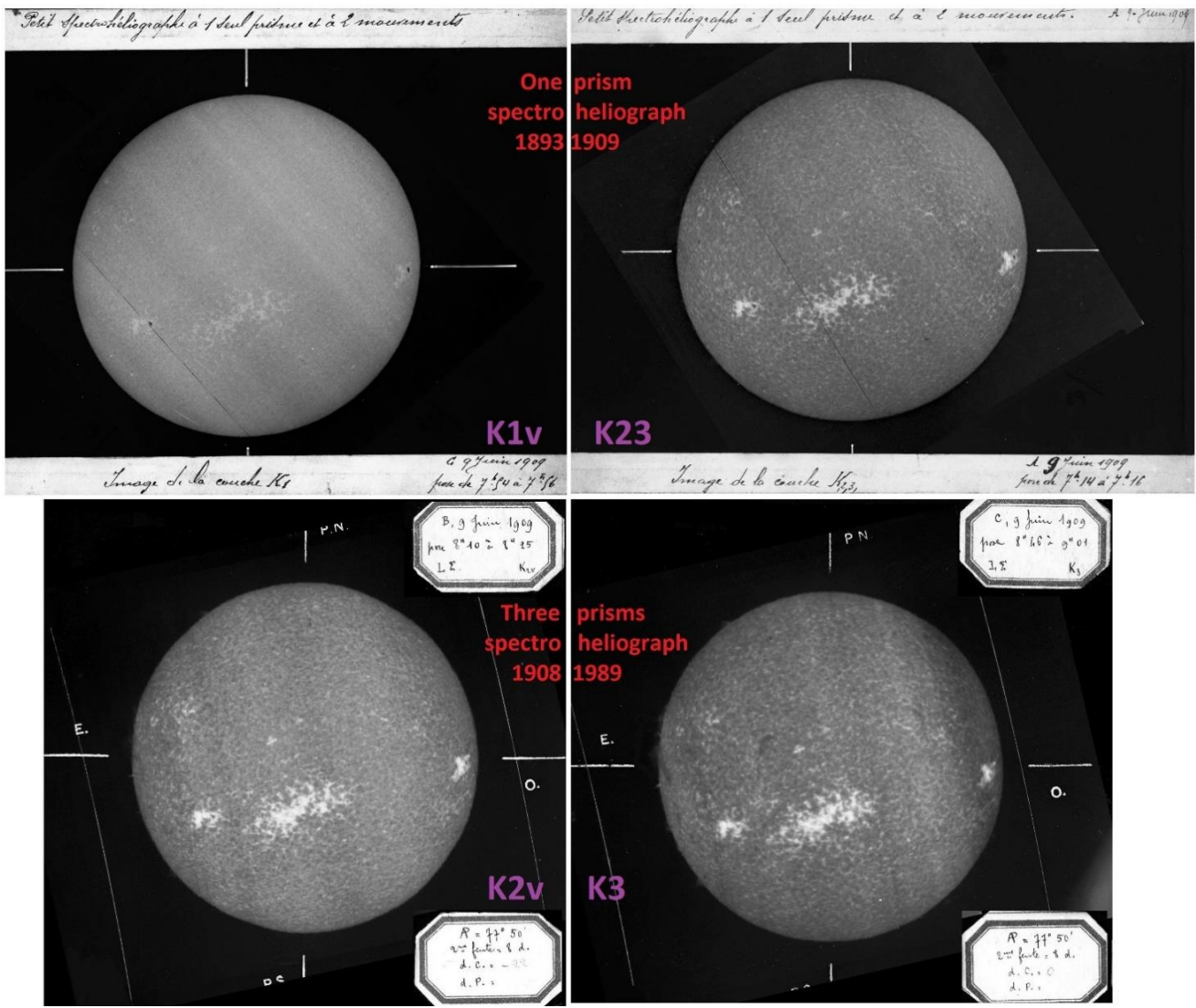

**Figure 25.** During a short period (1908-1920), the old spectroheliographs (Figure 14) and the new spectroheliographs (Figure 21) worked together, providing, on one hand, CaII K1v and CaII K23 images (top, single prism), and on the other hand, CaII K2v and CaII K3 images (bottom, three-prisms, narrow bandpass). Filaments appear in K3 only, because the K23 bandwidth is too large. 9 June 1909. Courtesy Paris observatory.

The multi-purpose 3.0 m spectroheliograph (3 of Table 2) and the high dispersion spectroheliograph (7.0 m or 14.0 m, respectively 4 or 4bis of Table 2) are shown in Figure 26. They were intensively used by L. d'Azambuja for his thesis work, that he defended in 1930. He carefully studied the dispersion of all optical combinations as a function of wavelength, either with prisms of gratings (Figure 27 and Table 3). The 7.0 m spectroheliograph (4) provided the best performance with the grating (2.0 Å/mm), but as a counterpart, the image size was very large (205 mm) upon the second slit in the spectrum, and a lot of scattered light was present. For these reasons, a second 7.0 m low dispersive spectrograph, with a third slit, was added (4bis). In this 2 x 7.0 m instrument, the scattered light was reduced by the second (enlarged) slit at the focus of the first spectrograph, and parasitic orders were eliminated by the single prism (o, Figure 26). The image reduction was in the ratio of the focal lengths of the collimator mirror (n, 7.0 m) and the chamber lens (called p, q, r, s by Deslandres, or $O_1$, $O_2$, $O_3$, $O_4$ by d'Azambuja, respectively 2.0, 1.3, 0.65, 0.30 m), providing the magnification factors of Table 2. D'Azambuja compared the capabilities of the spectroheliographs installed in various countries between 1892 and 1908. Table 4 indicates that Meudon spectrographs were among the best available instruments, and that the 7.0 m research spectroheliograph was probably unique.

(a) moving entrance objective (4 m focal length)
(T, M) speed converter + motor
(b) entrance slit
(c) collimator lens         (h) flat mirror
(d) grating                 (h') chamber lens
(g) three prisms            (k) output slit + moving photographic plate

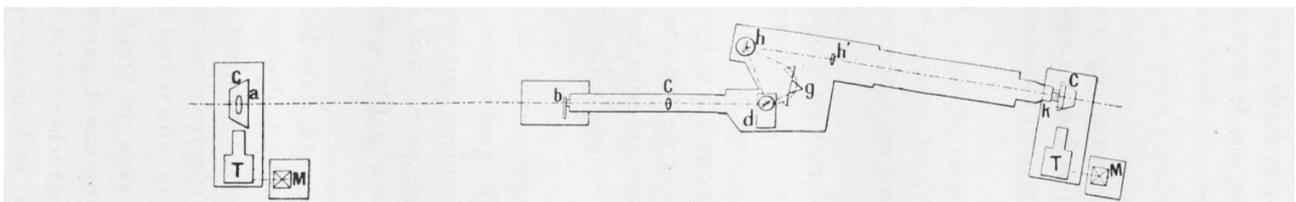

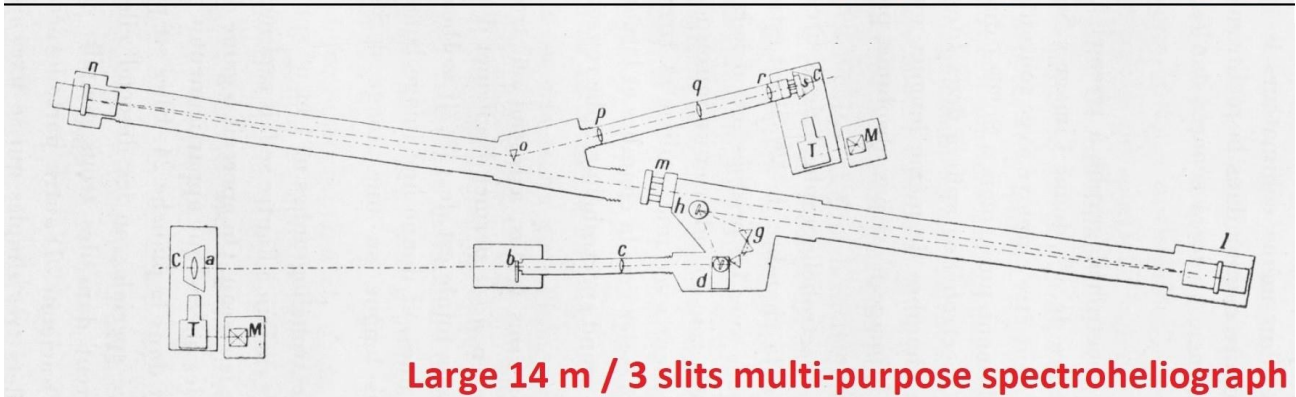

(a) moving objective              (h) flat mirror                                      **Image reduction (afocal system):**
(T, M) speed converter + motor    (l) concave mirror (7 m focal length chamber)        (n) concave mirror (7 m focal length)
(b) entrance slit                 (m) pre-slit in the spectrum (scattered light reduction)   (o) prism (order selection)
(c) collimator lens                   image diameter = 200 mm                          (p, q, r) lenses for image reduction
(d) grating                                                                                    (10 to 60 mm diameter)
(g) three prisms                                                                       (s) output slit and moving plate

**Figure 26.** The research spectroheliographs (3-4-4bis). The objective (a), entrance slit (b) and collimator (c) is common to all combinations, which could not be used simultaneously, but successively. Top: (3) is a multi-purpose, two-slit spectrograph with either three prisms (g) or a plane grating (d). Bottom: The large 14 m three-slit spectroheliograph (4 and 4bis) offers the highest dispersion and largest image. The chamber (right) is a concave mirror (l) of 7.0 m focal length, the spectrum is formed on the second slit (m). The diameter of the Sun is there 205 mm and the dispersion is the highest of all combinations (0.5-2 Å/mm depending on the wavelength). The second part (4bis, left) is optional: it is an afocal spectro-reducer using a second 7.0 m focal length concave mirror (n), a prism (o) to eliminate parasitic orders, a set of four reducing lenses (p, q, r, s) and a third output slit. This three-slit system reduced considerably the parasitic light. M and T were respectively electric motors and velocity transformers of the entrance objective and plate carriers. After Deslandres, 1910.

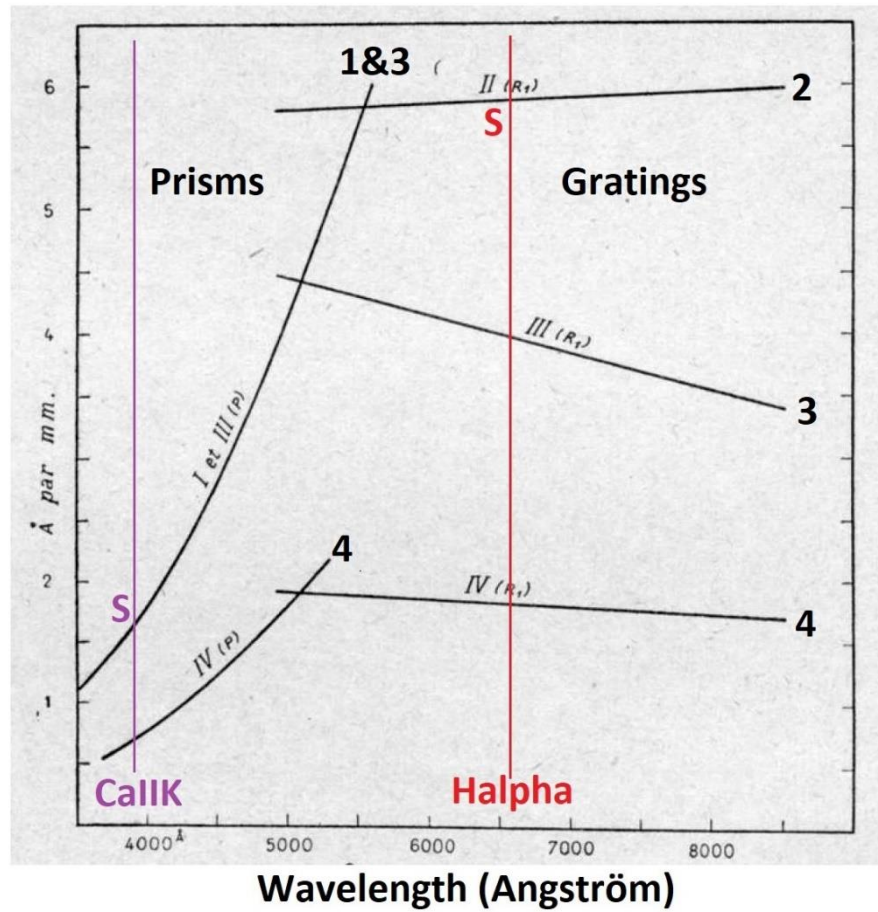

**Figure 27.** The dispersion of the quadruple spectroheliograph as a function of wavelength (in abscissa). For systematic observations (S), instruments 1 (I) and 2 (II) provided respectively 1.5 Å/mm for CaII K (three prisms) and 6 Å/mm for Hα (grating). For research work, the multi-purpose instrument 3 (III) and high dispersion instrument 4 (IV) offered respectively 1-4 Å/mm and 0.5-2 Å/mm in the waveband 3500-8500 Å, with either three prisms or gratings. After d'Azambuja, 1930.

**Table 3.** Characteristics of the four spectrographs in 1909. After d'Azambuja (1930).

| Spectrograph number | Spectrograph type | Focal length (m) | Spectral range (Å) | Dispersion (Å/mm) at λ (Å) with prisms or grating | | Image diameter (mm) |
|---|---|---|---|---|---|---|
| 1 | 3 prisms | 3.0 | 3900-4100 | 3934 | 1.7 | | 86.5 |
| 2 | grating | 3.0 | 3600-9000 | 3934 | | 5.7 | 86.5 |
| | | | | 5184 | | 5.8 | 87.5 |
| | | | | 6563 | | 5.9 | 89.5 |
| 3 | prisms/grating | 3.0 | 3600-9000 | 3934 | 1.7 | 4.8 | 86.5 |
| | | | | 5184 | 4.6 | 4.4 | 87.5 |
| | | | | 6563 | 9.4 | 4.0 | 89.5 |
| 4 | prisms/grating | 7.0 | 3600-9000 | 3934 | 0.7 | 2.0 | 206.0 |
| | | | | 5184 | 2.0 | 1.9 | 205.5 |
| | | | | 6563 | 4.1 | 1.8 | 205.0 |

**Table 4.** Comparison of spectroheliographs of various institutes observing the CaII K 3934 Å line in 1909. The quadruple spectroheliograph of Meudon (1908) had the best dispersion, allowing detailed observations of line profiles. After d'Azambuja (1930).

| Date | Observatory | Author | Disperser | Chamber (m) | Dispersion (Å/mm) |
|------|-------------|--------|-----------|-------------|-------------------|
| 1892 | Kenwood | Hale | grating order 4 | 1.0 | 3.7 |
| 1893 | Paris | Deslandres | 1 prism | 1.0 | 16.9 |
| 1903 | Kensington | Lockyer | 1 prism | 1.5 | 11.3 |
| 1903 | Yerkes | Hale, Ellerman | 2 prisms | 1.0 | 8.5 |
| 1906 | Kodaïkanal | Smith, Evershed | 2 prisms | 2.0 | 4.2 |
| 1906 | Tortosa | Cirera, Balcelli | 1 prism | 1.0 | 22.5 |
| 1906 | Mount Wilson | Hale | 2 prisms | 1.5 | 5.2 |
| 1906 | Postdam | Kempf | grating order 4 | 0.6 | 5.6 |
| 1907 | Catania | Ricco | 2 prisms | 0.6 | 11.3 |
| 1908 | Meudon | Deslandres | 3 prisms | 3.0 or 7.0 | 1.7 or 0.8 |

What can we conclude, in terms of solar physics, from this experimental phase (1892-1908) during which many series of monochromatic observations were performed and allowed to explore the solar atmosphere at various altitudes ? In summary, the following results were established:

- The upper layers of the chromosphere appear in the cores of CaII K or Hα lines, and exhibit structures, such as dark filaments, which are not present in the lower layers revealed by the wings of the CaII K line (K1 and even K2)
- Dark filaments of the upper layer require high dispersion instruments, able to select the K3 centre of the CaII K line; this is not necessary for bright plages and prominences at the limb, for which the K23 waveband is sufficient
- Images in the line wings, such as CaII K1v, show the photosphere (sunspots and faculae)
- While Hale, Ellerman and d'Azambuja thought that the dark filaments on the disk were the signature of bright prominences at the limb, Deslandres was not convinced by this association. The debate was closed later (1928), when the community officially considered the filaments and prominences to be two different views of the same object, either in absorption on the disk, or in emission at the limb.

## 5  L. D'AZAMBUJA'S RESEARCH WORK AND THE SYNOPTIC MAPS PROGRAM

In 1914, solar observations were interrupted by WW1. Deslandres and d'Azambuja were mobilized, and there was no more activity at Meudon. Observations restarted in 1919 and d'Azambuja completed simultaneously his studies at the University. In 1923, Alfred Pérot (1863-1925), a Meudon astronomer also professor at Ecole Polytechnique, introduced one of his assistants, Marguerite Roumens, to work on stellar spectra. She was finally hired in 1925 in the solar service at the death of Pérot, and she got married later (1935) with L. d'Azambuja. With her help, he observed in 1926-1927 many lines for his thesis, still unexplored (in comparison to the now well known Hα and CaII K) with the large 7.0 m spectroheliograph of Figure 26, in order to probe the atmospheric stratification and investigate the properties of solar structures. More than twelve lines were chosen, rather formed in the lower atmosphere:  CaI 4227 Å, SrI 4078 Å, MgI 3838 Å, FeI 4046, 4104, 4132, 4144, 4202, 4384 Å, MgI 5184 Å, NaD1 5890 Å, Hβ 4861 Å, and infrared lines of ionized Calcium. Many of them were studied in various regions, such as plages, active regions, quiet Sun, and the results were reported in his thesis (d'Azambuja,1930). He explored these lines in comparison to broader Fraunhofer lines and demonstrated that the initial choice made by Deslandres (Hα and CaII K lines) was well optimized for synthetizing solar activity phenomena, both for the photosphere and the chromosphere. For example, ionized CaII lines such as H (3968 Å) and K (3934 Å) in the violet part of the spectrum, and the infrared triplet (CaII 8498 Å, 8542 Å and 8662 Å) are deep and large. H and K provided similar results. Figure 28 shows two lines of the infrared triplet observed by d'Azambuja. The contrast of chromospheric structures, such as bright facular regions and dark filaments, are not better than in the K line centre. Another Calcium line, the neutral CaI 4227 Å (Figure 29) revealed sunspots and faculae in the photosphere, but the results were close to the blue wing (K1v) of the K line. Hence, it became obvious that observations of the CaII K line centre (K3) together with the blue wing (K1v) were optimized for a long-term collection of monochromatic images. Today, CaII 8542 Å is routinely observed with the SOLIS instrument at Kitt Peak (USA) in order to produce full disk magnetograms (i.e. images of magnetic fields) of the chromosphere via the Zeeman effect [note 6]. It has been discovered recently that the CaI 4227 Å line is the most linearly polarized line of the limb spectrum (also called the "second solar spectrum"); it allows to estimate the intensity of unresolved turbulent magnetic fields, which escape to the Zeeman effect, via the Hanle effect [note 7]. Since the seventies, the magnetographs (i.e. telescopes recording solar magnetic fields) use iron lines for systematic observations of the photospheric fields (Mount Wilson, Kitt Peak, and satellites such as SOHO/ESA/NASA and SDO/NASA). D'Azambuja, who studied many of these lines, could not imagine the future developments of spectroscopy. At his pioneering epoch, the magnetic nature of sunspots was just discovered by Hale (1908), and the law of sunspots polarity (the 22-year magnetic cycle) was just established by Hale & Nicholson (1925).

The Balmer series of Hydrogen was investigated several times by d'Azambuja (Figure 30), but the Hα line appeared with no doubt to be the best choice for active regions and filaments. These observations revealed decreasing contrasts of plages and filaments along the Balmer series, as mentioned and discussed by d'Azambuja (1938). For that reason, they were never done on a regular basis. The infrared HeI 10830 Å was also studied by d'Azambuja (1938). This was the first world-wide observation of this line, which was never systematically reproduced (Figure 31). Daily observations of HeI 10830 Å started much later (1974) at Kitt Peak Vacuum Telescope (USA), and systematic observations from space of HeII 304 Å (the ionized Helium) began onboard the SOHO/ESA/NASA mission in 1996 and continues today with the SDO/NASA satellite. This extreme ultraviolet line shows the hot filament channels around the cold core of neutral Helium and Hydrogen.

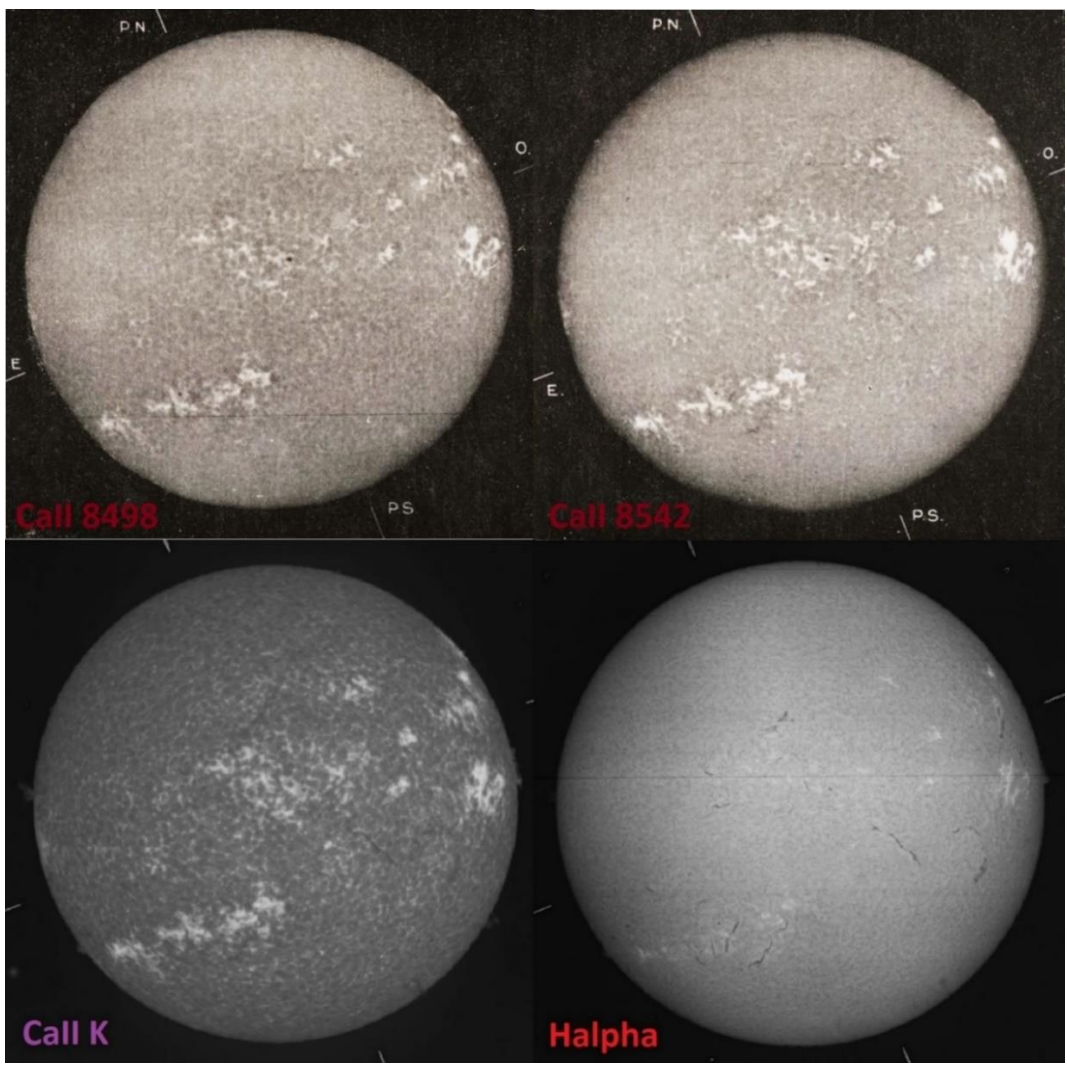

**Figure 28.** Test of ionized CaII infrared lines (8498 Å and 8542 Å, top), compared to the usual observations of CaII K and Hα lines (bottom). 9 September 1928. The structures appeared more contrasted in the K3 line centre (3934 Å) than in the infrared. We explain this result today by the formation altitudes of 1000 km, 1250 km (low chromosphere) and 1800 km (chromosphere) respectively for these three Calcium lines. After d'Azambuja (1930) and courtesy Paris observatory.

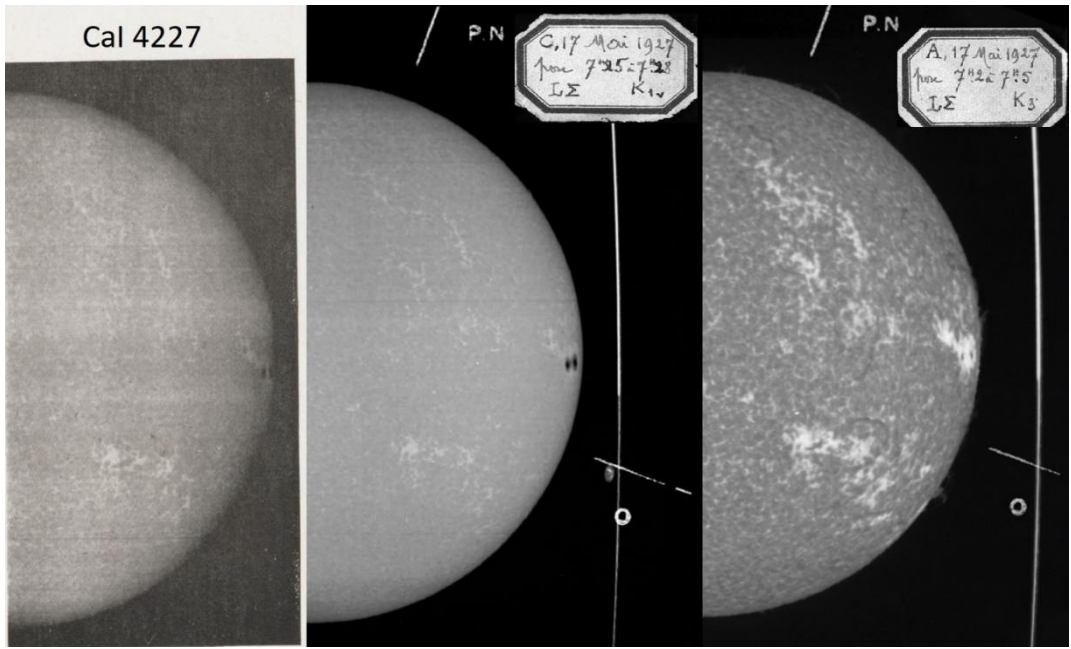

**Figure 29.** Test of neutral CaI 4227 Å line (left). Other wavelengths (right) are the daily spectroheliograms in CaII K1v (blue wing) and CaII K3 (line core). CaI 4227 Å provides images of the photosphere (sunspots, faculae) that are quite similar to CaII K1v images. 17 May 1927. After d'Azambuja (1930) and courtesy Paris observatory.

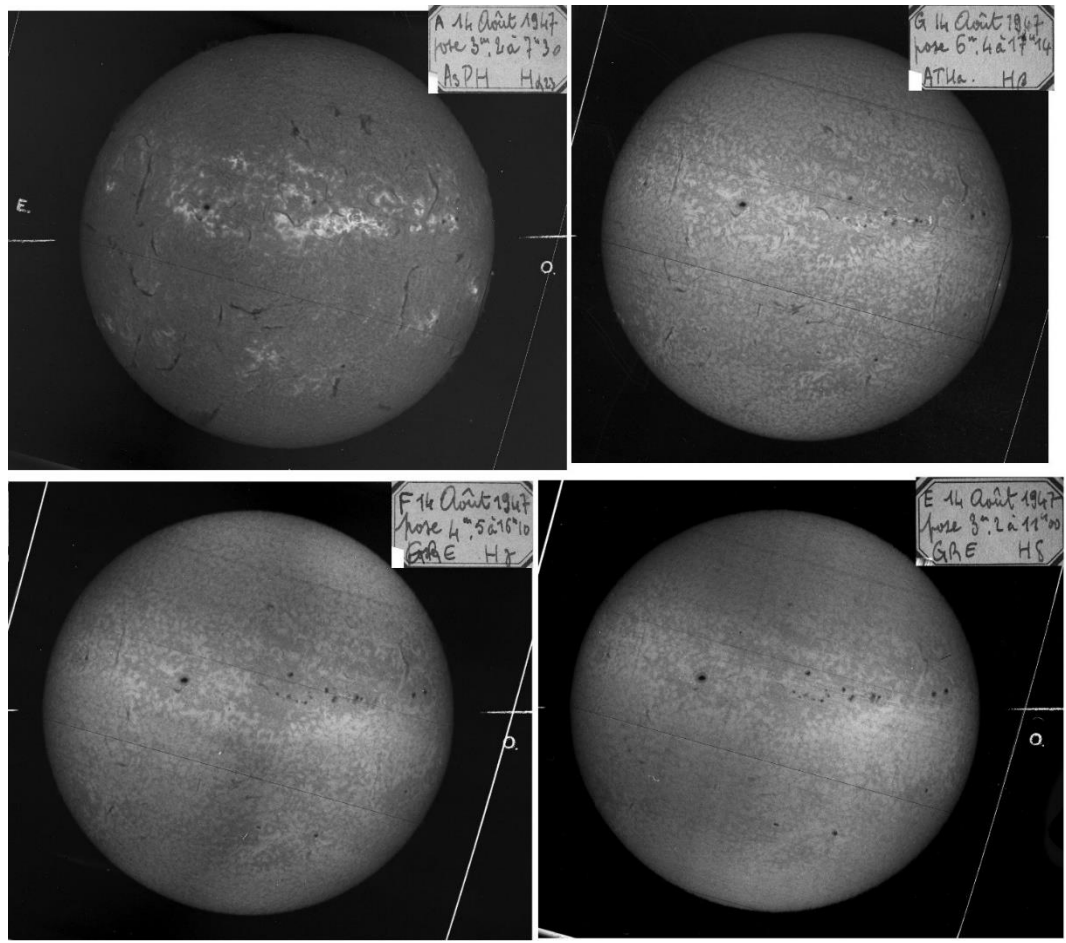

**Figure 30.** Test of the Balmer series Hα (6563 Å), Hβ (4861 Å), Hγ (4340 Å), Hδ (4101 Å). The contrast of solar structures, such as filaments and bright plages, decreases along the Balmer series. Lines become fainter and form deeper in the low chromosphere. Hε (3970 Å, not shown) was also observed, but it is located in the broad wing of the CaII H line (3968 Å) and impossible to isolate properly. 14 August 1947. Courtesy Paris observatory.

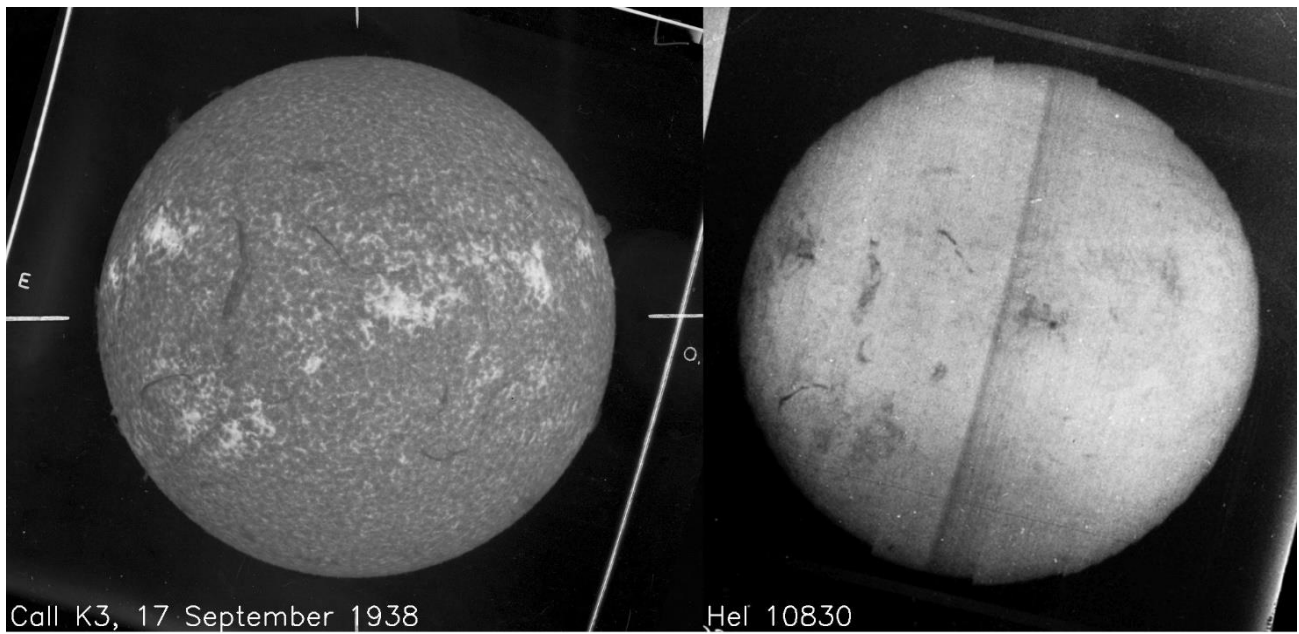

**Figure 31.** Test the infrared line HeI 10830 Å (right), the first world-wide observation. The usual CaII K3 image is displayed (left) for comparison. 17 September 1938. After d'Azambuja (1938) and courtesy Paris observatory.

Catalogs and drawings of the white light photosphere (sunspots, faculae) and prominences at the limb were undertaken in 1883 at Zürich observatory ("*Eidgenössische Sternwarte*", Switzerland, Illarionov & Arlt, 2022). In order to study the long-term solar activity and, in particular, the behaviour of solar filaments, plages and sunspots, together with their mutual influence, d'Azambuja took advantage of multi-layer monochromatic images. He started in 1913 a more ambitious program, the drawing of synoptic charts of mixed structures of the upper layer (filaments of the chromosphere) and the lower layer (sunspots). A preliminary map was published after WW1 (d'Azambuja, 1921) and presented at the IAU colloquium in 1922. At the 1925 symposium, 30 rotations were shown, and an international diffusion of the Meudon maps was proposed. The CaII K3 and Hα spectroheliograms were used respectively for bright plages and filaments, and CaII K1v for sunspots. It consisted in reporting, for each synodic rotation of the Sun (mean duration of 27.2753 days), the average position of solar structures on rectangular maps, with the longitude in abscissa and the latitude in ordinates. Each rotation is identified by a Carrington number starting on 9 November 1853. The collection begins at rotation 876 (March/April 1919, Figure 32) and ends at rotation 2008 (October 2003). Each map is accompanied by various tables listing many parameters, such as coordinates, lifetime, length, height or area of the observed structures (Figure 33).

When Meudon data were lacking because of clouds, an agreement with Kodaïkanal (India), Mount Wilson (USA) and Coïmbra (Portugal) was established to use their respective images. The synoptic charts were periodically published by d'Azambuja (1928) in the Annales of the Paris-Meudon observatory, and later in l'Astronomie. M. d'Azambuja contributed a lot to this long-term undertaking which implied also G. Olivieri and R. Servajean. The conversion of the spherical coordinates of spectroheliograms into rectangular ones was a complicated and boring task. For that reason, R. Servajean and H. Grenat (1900-1968) imagined in 1949 an anamorphic device (Figure 33), in order to optically convert spectroheliograms into planispheres. This surprising instrument was built by Marcel Brebion, a technician in mechanics, who received the diploma of "*un des meilleurs ouvriers de France*" (one of the best workers of France). Olivieri wrote, in his memoirs, that "*Mrs d'Azambuja was extremely severe concerning the quality of the work, and that he succeeded sometimes to produce results which were accepted by Mr d'Azambuja, but which were previously refused by his wife*"! Mrs d'Azambuja had indeed a so high regard for the reputation of the Meudon group that Milosevic, a visiting astronomer, said someday this humoristic remark: "*it is as if the Sun was the property of Mrs d'Azambuja*".

The synoptic maps were the basis of the monumental research memoir about solar filaments published by d'Azambuja & d'Azambuja (1948). It was considered like a "*bible*" among Meudon astronomers and is quite comparable to a thesis. The morphological properties of filaments were investigated, as well as their rotation and motions, together with their relationships with surrounding sunspots and plages. After the retirement of M. d'Azambuja in 1959, the synoptic map program continued under the auspices of M.-J. Martres, who was hired in 1955 and in turn retired in 1988. She published in l'Astronomie the maps of the period 1958-1991 (more than 400 rotations) and many scientific papers based on the interpretation of synoptic charts concerning filaments and solar activity. The drawings were unfortunately stopped in 2003 when the last technician (Germaine Zlicaric) left Meudon. Existing maps are available at ftp://ftpbass2000.obspm.fr/pub/synoptic/ and contain probably still unexplored information concerning solar activity.

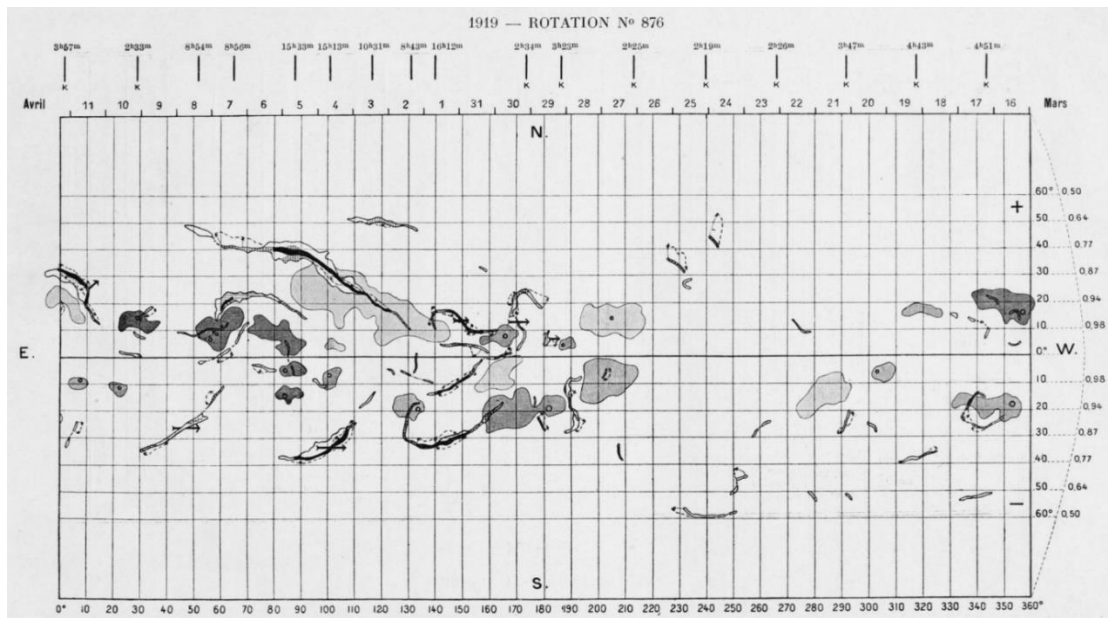

**Figure 32.** The first synoptic chart of the chromosphere and sunspots started at solar rotation 876, March/April 1919 (abscissa: longitude; ordinates: latitude). The maps represent, for each rotation, sunspots, filaments and bright plages and are accompanied by tables (Figure 33) characterizing the observed structures. After d'Azambuja (1928).

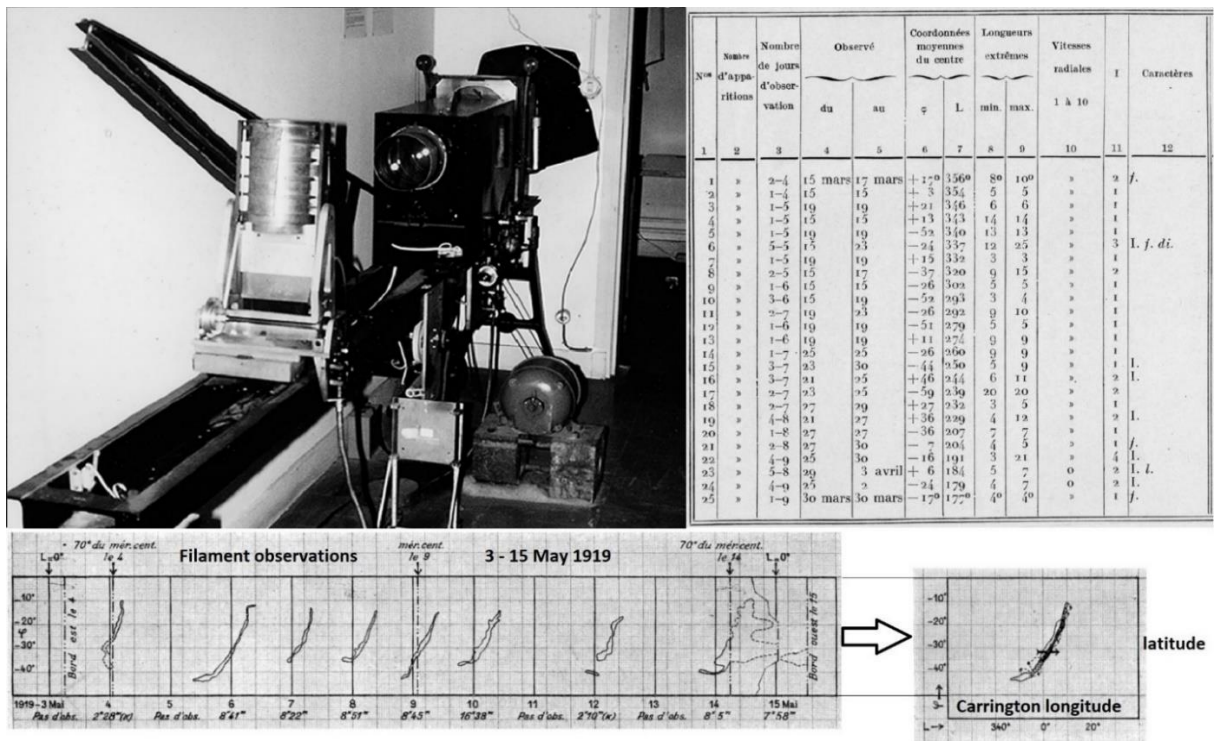

**Figure 33.** L. d'Azambuja initiated the edition of synoptic maps of the chromosphere and sunspots. Each map was supplemented by a table. The first published one (right, associated to the rotation 876 of Figure 32) concerns filaments and provides their observing dates, coordinates, lengths, and indices reporting radial velocities (range 0-10, unit = 5 km/s) deduced from section spectroheliograms (when available) and significance (range 1-10). The last column contains comments (S=stable, I=unstable, f=thin, l=thick, di=discontinuous…). The bottom panel explains the drawing method of filaments: the envelope of the varying structure is reported on the synoptic map. The anamorphic optical system (left) was imagined by Servajean to simplify the process. Courtesy Paris observatory and after d'Azambuja (1928).

## 6 SPECIFIC SPECTROHELIOGRAMS FOR RADIAL VELOCITIES (1919-1939)

Section spectroheliograms were made of contiguous spectra of the CaII K line, recorded step by step via discontinuous motions of the entrance objective and the photographic plate. The spectra were cross sections of the Sun (approximatively North/South); depending on the spatial step of the sections, or on the waveband of each section fixed by the output slit, the final image on the photographic plate was either elliptic (Figure 34, step 5") or circular (Figure 35, step 22"). The line profiles were used to determine the motions of the structures from the Dopplershift of the central component (K3), as shown by Figure 34 (right). The K3 shift (named α in the figure) is proportional to the line-of-sight (or radial) velocity; it was measured from a known wavelength origin. Such measurements were useful to investigate the dynamics of filaments and were reported in the catalogues associated to synoptic maps, via an index (0-10, unit = 5 km/s) and a sign (blue/red shift).

Figure 35 displays a section spectroheliogram, when observations were organized on a regular basis after WW1, together with the standard monochromatic images (86 mm diameter) obtained daily in CaII K1v (photosphere), CaII K3 and Hα (chromosphere). The number and waveband of cross sections, and the scanning step of the Sun, were calculated to produce circular images, with about 86 sections of 1 mm (corresponding to the waveband of 2.0 Å and the step of 22" on the Sun), contiguously recorded on the photographic plate. Such observations of CaII K line occurred two times daily (one for the full disk and another one for remarkable prominences), with the priority to the usual spectroheliograms. For instance, in 1920, 333 observations with the section spectroheliograph were performed (providing a total of 1200 images with monochromatic images). Deslandres (1924) explains that daily observations consist of: *"(i) the lower layer in CaII K1v for sunspots and faculae; (ii) the upper layer in CaII K3 for dark filaments and prominences; (iii) the upper layer in Hα, revealing more details; (iv) two observations for radial velocities, one for the disk and another for prominences"*. However, the analysis of line profiles and the velocity measurements were so delicate and so time consuming for the operators (which were not numerous and also observers) that this specific observation was abandoned in 1939, at the entry of WW2, and never restarted after. But the monochromatic observations of the photosphere (CaII K1v) and the chromosphere (CaII K3, Hα) were fortunately preserved during the war.

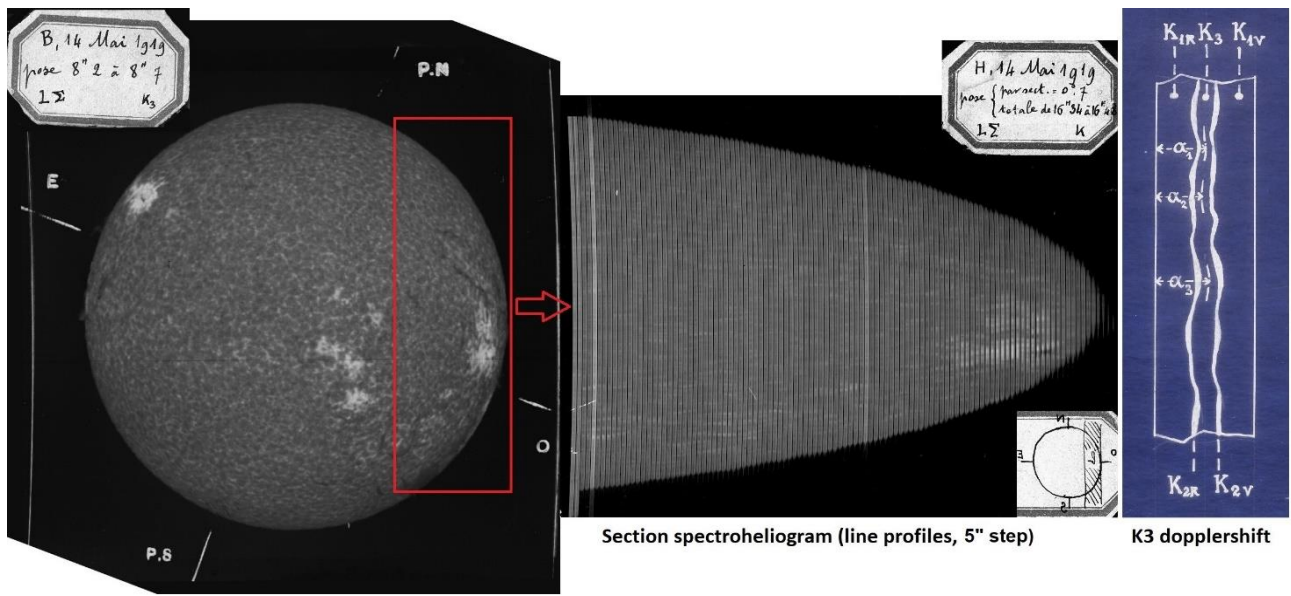

**Figure 34.** Observations for the measurement of radial (line-of-sight) velocities, 14 May 1919. The centred image is a section spectroheliogram of the western part of the Sun (the rectangle reported on the monochromatic CaII K3 image). It is made of the juxtaposition of spectra (5" step) of the CaII K line. The drawing (right) shows how was determined the Dopplershift α of the line core (the K3 component). Courtesy Paris observatory and d'Azambuja (1920b).

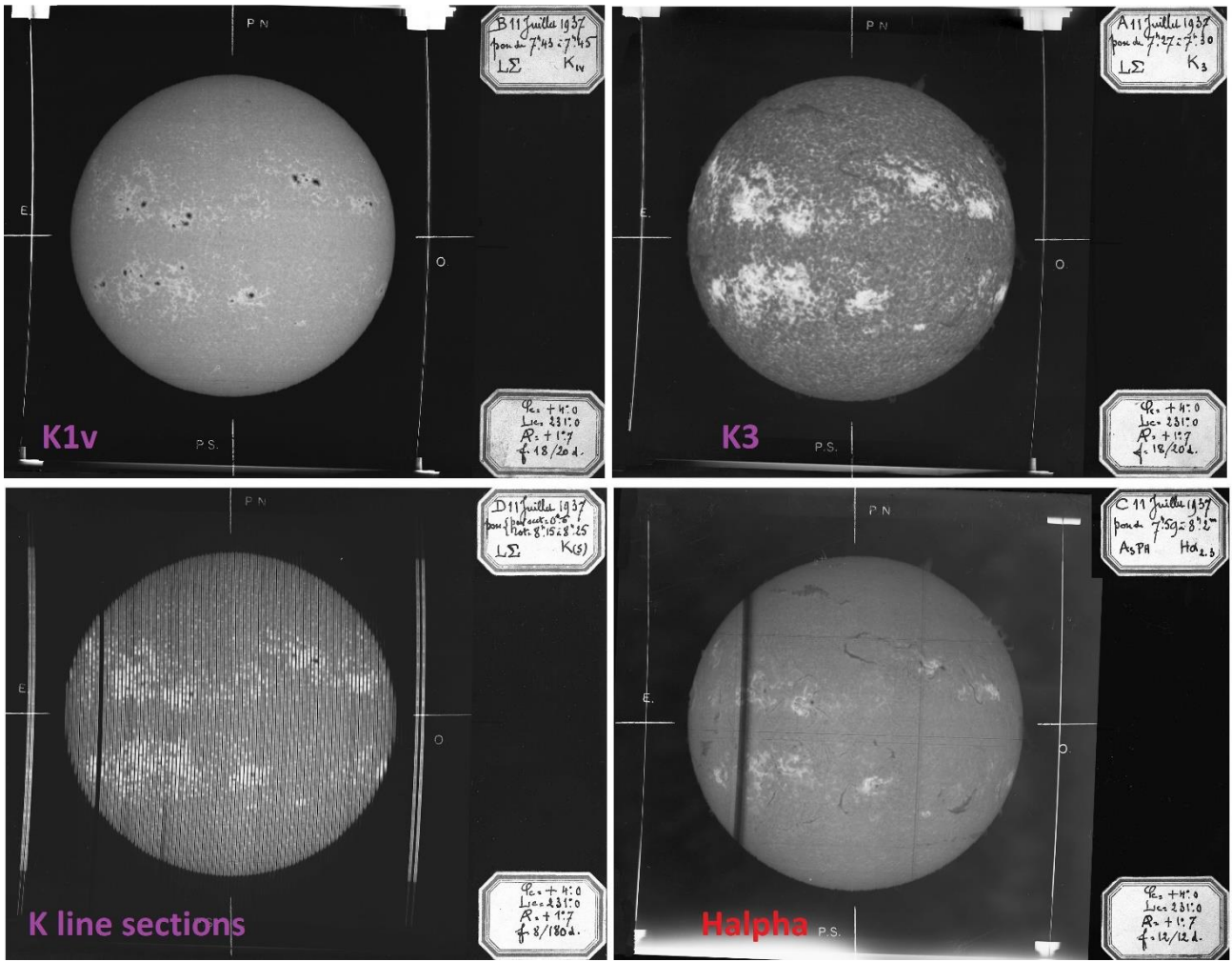

**Figure 35.** Standard observations made daily until 1939 (example of 11 July 1937). Top: CaII K1v (photosphere) and CaII K3 (chromosphere). Bottom: the section spectroheliogram for radial velocity measurements (CaII K line profiles, 22" step) and the usual Hα image. Courtesy Paris observatory.

**7 THE GOLDEN AGE OF SOLAR SURVEY AND THE INTERNATIONAL ORGANIZATION AFTER WW1**

We saw that systematic observations were totally interrupted by WW1. They resumed in 1919 with the two optical configurations of 1909 (Figure 22), corresponding to the pictures of Figure 36, taken in 1921 a few years after WW1. The CaII K line is dispersed by three prisms; the corresponding chamber (A2) is located at the left of the observer who stands just besides the 37 mm solar image and the entrance slit (S) of the spectrograph. The Hα line is formed by a plane grating; the corresponding chamber (A1) is above the right pillar. The carriers of photographic plates of (A1, A2) are visible, together with the output slits (75 µm) of the chambers, which select the lines. The diameter of monochromatic images is 86 mm. It is interesting to notice the evolution since 1909 (Figure 21): the wooden supports were replaced by brick pillars and a new mechanical device (N) was suspended to the ceiling. (N) is still running today: it is a mechanical coupling between the motor (M) which translates the imaging objective (not visible) and an artificial "moon" in the image plane. It is made of a round neutral density 1 (ND1, 37 mm diameter, 10% transmission) moving with the solar image during the surface scan, in order to observe faint structures at the limb. Indeed, exposure times required for prominences are longer than for the solar disk; the ND1 mask allows to observe them without any disk burning (apart the chromosphere just at the limb). As the angular size of the Sun varies with seasons, two ND1 attenuators with different diameters are available, a "winter moon" and a "summer moon".

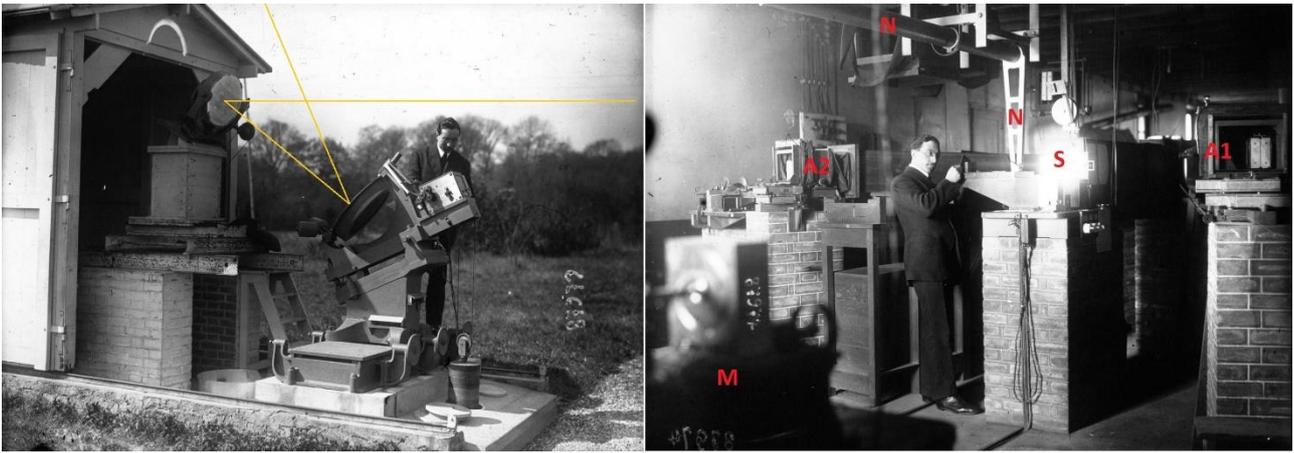

**Figure 36.:** the two-mirror coelostat catching the Sun in 1921 (left). After the two reflections, the horizontal beam crosses the objective (not shown) to form a 37 mm solar image upon the spectrograph slit (S, right). It moves owing to the motor (M) of the objective. The CaII K (A2) and Hα (A1) chambers have motorized photographic plate supports, which move 2.31 times faster than the objective. Inside the (A1) chamber, one sees the output slit. (N) is a metallic rotating pipe which is mechanically coupled to the carrier of the imaging lens and drives an attenuator in the solar image, allowing to observe both the disk and limb structures with long exposure time. Courtesy Gallica/BNF.

In agreement with Deslandres's formula ("*one must record continuously all variable phenomena of the Sun and Earth*" to understand their mutual interaction), the spectroheliograph was supplemented in May 1921 by a magnetometer to follow the fluctuations of the Earth's magnetic field. It was located in the vicinity of the solar instruments and made of a magnetized needle and recorder, in order to study the influence of sunspots and plages upon the terrestrial magnetic field direction. When the needle deviation was important, it triggered a ring to alert the solar observers that something was happening.

Systematic observations were not interrupted by WW2, apart the the section spectroheliograms of CaII K line which were definitely stopped in 1939. After the war, André Danjon (1890-1967), who was the director of Paris-Meudon observatories from 1945 to 1963, organized three scientific teams: "the spectroheliographs" (daily solar observations of the photosphere and the chromosphere), "the external layers of the Sun" (prominences and hot corona), "the comets and nebulae", respectively headed by L. d'Azambuja, B. Lyot and F. Baldet (Figure 37). At the retirement of L. d'Azambuja in 1954, M. d'Azambuja became responsible of the spectroheliograms and synoptic maps. B. Lyot dramatically passed away in 1952 during the Khartoum eclipse mission, and was replaced in 1954 by Raymond Michard (1925-2015) at the head of the "external layers" team. At the retirement of M. d'Azambuja in 1959, Michard became responsible of a unique solar service, which was transformed into the solar and planetary department (named DASOP) after the events of May 1968, according to the new status of the institution (Mein & Mein, 2020). Michard stayed at the head of DASOP until 1971, and became the first elected President of Paris observatory (including Meudon and the Nançay station).

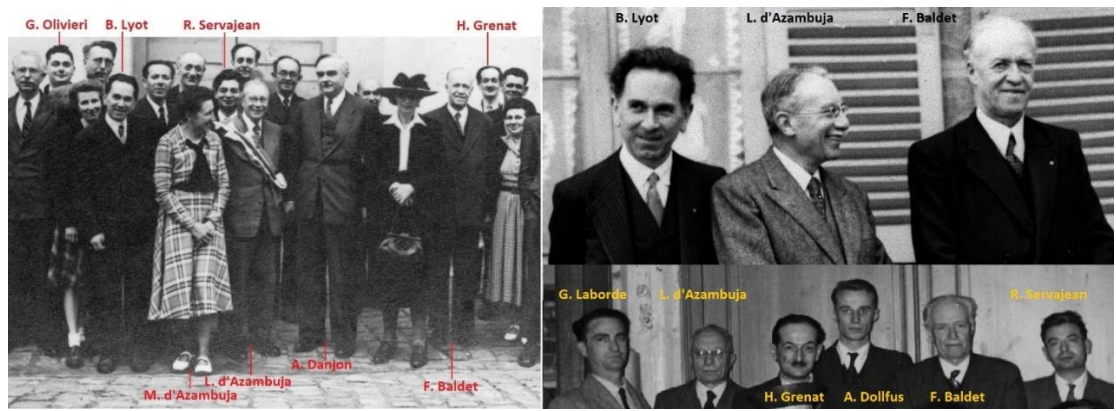

**Figure 37.** The Meudon scientific staff in the fifties. Left: L. d'Azambuja's jubilee, after fifty years of astronomy at Meudon (1949). Right (top): in 1950, Lyot, d'Azambuja and Baldet were at the head of the three scientific teams, respectively the services of spectroheliographs, external layers of the Sun, comets and nebulae. Right (bottom): Laborde, Grenat and Servajean were affected to d'Azambuja's service. Courtesy Paris observatory.

In order to follow in real time the solar activity, Hale (1929) invented the visual spectrohelioscope. It was a spectroheliograph with two oscillating slits (Ré, 2014), or much better, two Anderson's prisms (of square section) rotating at the same speed in front of the input and output slits. In that case, the first prism (at the entrance slit in the solar image) was used to scan the Sun and the second prism (behind the slit in the spectrum) formed a monochromatic image due to the retinal persistence. The international cooperation for solar observations began in the thirties. Hale (1931a) was a precursor and initiated a network of 23 visual spectrohelioscopes, which were identical and disseminated in many stations to establish a world-wide survey of chromospheric activity. In 1932, the solar commission of the IAU split in four groups and d'Azambuja became responsible of the subdivision dedicated to chromospheric phenomena. D'Azambuja (1939) described the organization for the coordination of the 23 stations located in America, Europe, Asia, Australia, New Zeeland. The observations of flares were collected and compiled by Meudon, which published lists, indices and tables in the Bulletin for Character Figures of Solar Phenomena, edited by Zürich observatory since 1928. It was renamed Quarterly Bulletin on Solar Activity (QBSA) in 1939 and was transferred in 1976 to Mitaka (Japan, archive at https://solarwww.mtk.nao.ac.jp/en/wdc/qbsa.html). The QBSA was delivered in 60 countries until 2009.

Many results were obtained with the spectrohelioscopes (Hale, 1930, 1931a, 1931b). At Meudon, it was possible to switch quickly between the photographic spectroheliograph and the removable visual instrument of Figure 38 and Figure 39. It was based on Anderson's rotating prisms. The spectrohelioscope was used daily during two hours for inspection of active regions at the surface of the Sun, in order to detect fast evolving or dynamic events, such as flares or filament instabilities (after 1955, this task was transferred to monochromatic heliographs using Lyot filters). In the fifties and during the summer, the director of Paris-Meudon observatories, A. Danjon, lived in Meudon. His main entertainments were picking mushrooms in the forest and observations with the visual spectrohelioscope. This device is still present but no longer active. It must be noticed that a spectroheliokinematograph was built at Kitt Peak by McMath & Petrie (1934) using 35 mm motion picture cameras to record high cadence observations, but this invention was soon supplanted by much more convenient telescopes using filters (Lyot, 1944). The optical survey of solar activity was supplemented after 1948 by radio wavelengths. Laffineur & Houtgast (1949) transformed a german Würzburg radar of WW2 (7.50 m diameter) installed at Meudon into a radiotelescope and detected solar bursts at the frequency of 550 MHz (55 cm wavelength). This antenna is the ancestor of the Nançay instruments (after 1953).

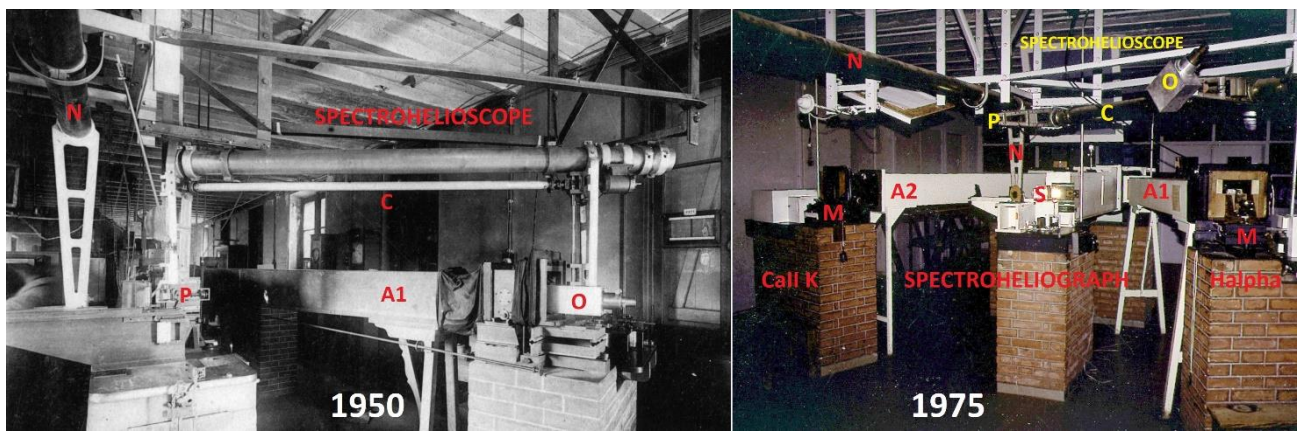

**Figure 38.** Left: the visual spectrohelioscope in the fifties worked with two fast rotating prisms, the first one in front of the entrance slit (P) and the second one behind the second slit in the spectrum (O). There was a single motor for both prisms with a mechanical coupling (C). An ocular in (O) allowed to see the exit slit, which formed a spectroheliogram due to light persistence on the eye's retina. The (A1) chamber only (for Hα images) was concerned. Right: the photographic spectroheliograph (A1 for Hα, A2 for CaII K) and the visual spectrohelioscope (not in observing position), 25 years later, in 1975. Courtesy Paris Observatory.

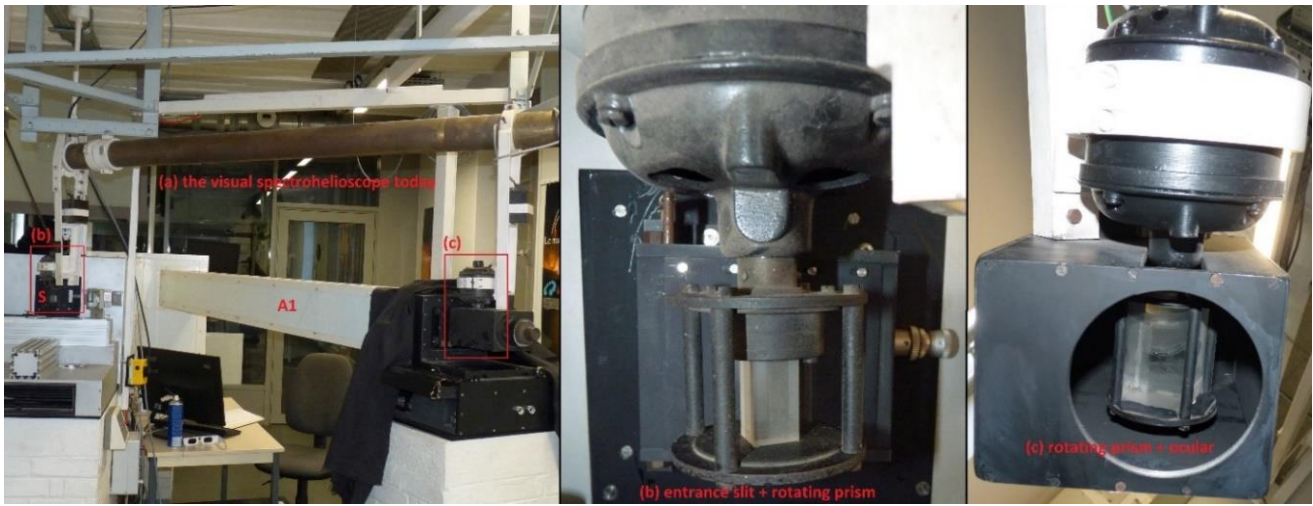

**Figure 39.** (a): the visual spectrohelioscope worked with two high speed rotating prisms, respectively at the entrance slit (b box) and at the output slit (c box) in the spectrum. Two synchronous motors replaced the single motor of Figure 38, so that the complex mechanics was removed. (b): the prism (14 mm square section) in the image plane (motor above). (c): the prism (32 mm square section) in the spectrum between the exit slit and the ocular (motor above). The section ratio (32/14) corresponds to the spectrograph magnification (2.31). Courtesy Paris Observatory.

In order to investigate the impact of solar flares at the Earth, such as ionospheric perturbations and events driven by the solar wind, occurring generally within a 24-hour delay, messages (called ursigrams) were broadcasted to the geophysical community by several countries under the auspices of the "Union Radio Scientifique Internationale" (URSI), as early as 1928. An index (0-5) provided the solar activity level. The broadcast was interrupted by WW2, but the IAU recommended the Commission 11 (presided by L. d'Azambuja) to examine the question of "*the prediction of terrestrial phenomena of solar origin and the determination of active areas*" (d'Azambuja, 1949). Ursigrams restarted in June 1947 from information collected by many observatories and sent to Meudon by telegram. It was not possible to predict solar flares, but the messages indicated daily the position of active regions and the occurrence of flares. As this tool was very important for studies of solar terrestrial relations (called Space Weather today), it was extended and new broadcasting centres appeared. In 1962, the International Ursigram and World Days Service (IUWDS) was created and favoured the development of solar activity forecasting methods. It was replaced in 1996 by the International Space Environment Service (ISES, http://www.spaceweather.org/), with 15 Regional Warning Centres (RWC) at various longitudes (Meudon was RWC until 1997). Among them, the Space Weather Prediction Centre (Boulder, USA, https://www.swpc.noaa.gov/) emits daily forecasts and is nowadays the most exhaustive service. The Scientific Committee on Solar Terrestrial Physics (SCOSTEP, https://scostep.org/) coordinates international research programs to improve our understanding of solar instabilities and the forecasting reliability. This implies also international campaigns of multi-wavelength observations (space for high energy radiations, ground-based for optical and radio events), such as the Max Millennium program (https://solar.physics.montana.edu/max_millennium/).

At the retirement of Mrs d'Azambuja (1959), observations continued at Meudon with Servajean, Martres and Olivieri, who left the observatory respectively in 1975, 1988 and 1992. Mrs d'Azambuja was also strongly implied in synoptic maps and in the flare catalogue of the QBSA. She was also involved in the preparation of the French program of the International Geophysical Year (IGY, 1957), together with Michard. He initiated new instruments dedicated to solar activity studies (such as the flare spectrograph at Pic du Midi or high cadence monochromatic heliographs at Meudon and Haute Provence observatories, see Mein & Mein, 2020).

After the d'Azambuja's retirement (1954/59 respectively for Mr and Mrs), the very powerful 7.0 m spectroheliograph of Figure 26 (mainly used for research purpose) was slightly transformed by Michard & Rayrole (1965) to produce, as often as possible, magnetic field plates of sunspots and active regions (PCM, in French "*Plaques de Champs Magnétiques*", Figure 40). About 100 observations were done in a few years. They used a Hale-Nicholson grid located at the entrance slit of the spectrograph fed by the 75 cm Foucault Sidérostat (located at right of Figure 19) and a new 40 cm (25 m equivalent focal length) gregorian telescope providing a 235 mm solar image. However, it worked exactly like the section spectroheliograph of Deslandres, forming spectra of an iron line (sensitive to the Zeeman effect [note 6]), and scanning active regions step by step (the telescope and the plate were moving simultaneously). The dispersion of the large 7.0 m spectrograph was 3.85 mm/Å for the FeI 6302.5 Å line, owing to the new 1200 groves/mm Baush & Lomb grating. Each PCM (9 x 24 cm²) had a spatial step of 2.4" or 3.6" and contained a maximum of 110 sections of 2 mm, each having a spectral width of 0.52 Å. The grid was composed of circular polarizing strips (alternatively quarter and three-quarter wave) providing the σ+ and σ- signals of the Zeeman effect. An observer (retired a long time ago) told me that the first grids were made of the cellophane (a birefringent material) composing the stocking packages of Mrs Michard ! For FeI 6302.5 Å, the Zeeman splitting (w = 0.46 B, in Å) is proportional to the magnetic field B (in Tesla). The wavelength shift between σ+ and σ- is (2 w). In sunspots (B ≈ 0.1 Tesla), the order of magnitude of the Zeeman shift is 0.1 Å. This beautiful experiment, based on d'Azambuja's spectrograph, was the ancestor of the Meudon magnetograph built by Jean Rayrole (1932-2008) in the eighties and the THEMIS telescope (Tenerife), designed by Rayrole and the French INSU/CNRS institute in the nineties.

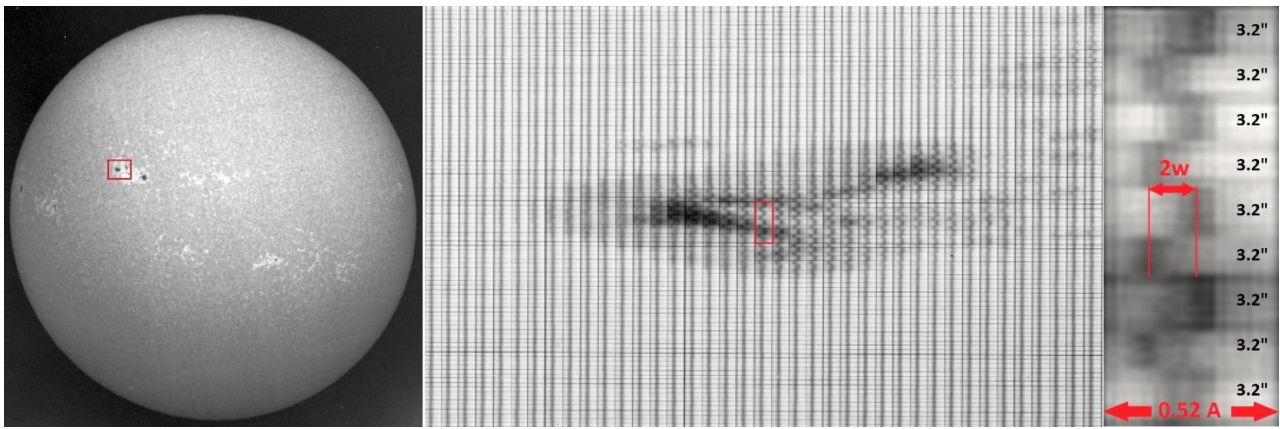

**Figure 40.** Magnetic fields studied with d'Azambuja's 7.0 m spectrograph. Left: CaII K1v spectroheliogram of 10 June 1963; the red box corresponds to the field of view of the magnetic field plate (PCM). Centre: the PCM, observation of FeI 6302.5 Å in sunspots with the Hale-Nicholson grid, by spatial steps of 2.45" in the East/West direction. Horizontal strips of 3.2" height provide alternatively σ+ and σ- signals along the North/South directed slit. The red box corresponds to the magnified detail. Right: the line-of-sight magnetic field is proportional to the Zeeman shift (2 w) between σ+ and σ-. On this example, it indicates a magnetic field of 0.14 Tesla. Courtesy Paris observatory.

## 8 RARE EVENTS OF SOLAR ACTIVITY OBSERVED WITH THE SPECTROHELIOGRAPHS

The source of solar activity is located in the photosphere and in the chromosphere. Many spectacular events, such as instabilities of prominences of Figure 41, were often observed in the past with the spectroheliographs, both in CaII K3 and Hα. They showed that violent mass ejections (tens of km/s) often occur during active phenomena. D'Azambuja & Grenat (1926) described the solar flare of 13 October 1926 where velocities of 130 km/s were measured in Hα, and even 150 km/s in the yellow HeI D3 line (5876 Å), with the research spectroheliographs. The day after, a strong geomagnetic storm was detected (this is a perturbation of the Earth's magnetic field due to solar energetic particles, carried by the solar wind at 1500 km/s, and impacting for instance the telecommunications or electricity transport at high latitudes). A polar aurora (light emission related to the interaction between solar particles and the Earth's atmosphere) was also observed at Meudon observatory.

Before the fifties, high cadence telescopes using narrow filters did not exist, and spectroheliographs were the unique tool, although unable to sustain the fast observing cadence needed to follow the evolution of flares in details. Much faster instruments were developed for and after IGY 1957 with Lyot and Fabry-Pérot filters, mainly in Hα. Meudon used the first Lyot filter in 1954 with a 60 s cadence. Today, while ground-based instruments see the birth and the development of events in the chromosphere (8000 K), telescopes onboard satellites, such as SDO/NASA, observe ejections above the chromosphere, in the low corona, in the temperature range $10^5$-$10^7$ K owing to the extreme ultraviolet lines (only visible from space). White light coronagraphs, such as those onboard SOHO/ESA/NASA or STEREO/NASA, even permit to follow the propagation of mass ejections at large distances up to 30 solar radii.

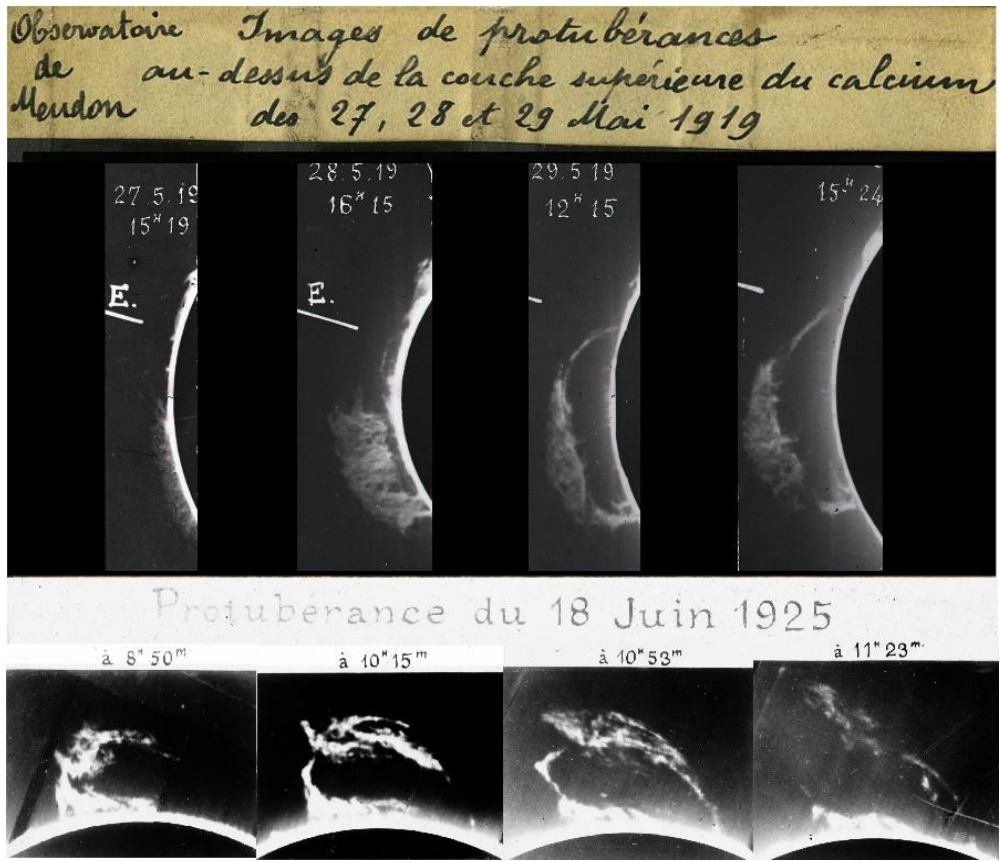

**Figure 41.** Instabilities of prominences at the limb observed in CaII K3 on 27, 28 and 29 May 1919 (top) and 18 June 1925 (bottom) with the spectroheliograph. The times are indicated in the picture. Courtesy Paris observatory.

The centennial Meudon collection of CaII K1v spectroheliograms contains plenty of information about the area covered by sunspots. In particular, the April 1947 sunspot group (Figure 42) is probably the largest ever seen (6.1% of the solar disk), followed by other groups of the same solar cycle (number 18), among them the February 1946, May 1951, July 1946 and March 1947 groups (respectively 5.2%, 4.9%, 4.7% and 4.6% of the disk). The July 1946 group (Figure 42) produced the strong solar flare of Figure 43, and a huge geomagnetic storm. This event is one of the most energetic eruptions ever observed. A. Danjon, the director of Paris-Meudon observatories, was there during this event. Olivieri (1993) reports in his memoirs that "*his advices and commands given to the observers did not helped them to stay quiet and keep cool*" ! Flares originate in active regions and have a magnetic origin. The magnetic energy is stored in sunspot groups; it is proportional to the surface and to the square of the magnetic field intensity. Unstable configurations often occur near the solar maximum and trigger huge flares. The 1946 event illustrates the importance of historical data. It allowed Aulanier *et al.* (2013) to predict the maximum energy of a solar flare, about $5 \cdot 10^{26}$ Joule (10 times the one ever measured by satellites of the modern era, the 4 November 2003). The average energy of solar eruptions lies around $10^{25}$ Joule; this is higher than the one consumed by the humanity during years ! But this is smaller than energies of super-flares ($10^{27}$-$10^{29}$ Joule) discovered by the Kepler satellite in Sun-like stars, which are suspected to exhibit much bigger spots than our Sun (Schmieder, 2018).

Figure 42 displays incredibly long solar filaments (more than a solar radius of 700 000 km, a rare phenomenon). They delineate giant cells of magnetic opposite polarities. The figure shows together bright regions (the facular areas) which were used by Chatzistergos *et al.* (2020) as a proxy to reconstruct solar magnetic fields over the full twentieth century, in particular before direct measurements of magnetographs (using the Zeeman effect), which appeared only in the seventies. This proxy allowed also Chatzistergos *et al.* (2021) to develop a method for the long-term determination of the Total Solar Irradiance (TSI or solar flux in Watt/m² at the Earth orbit); it is a promising tool to reconstruct the TSI since 1900 or earlier, which is of great interest to evaluate the solar influence on the Earth's climate, at epochs when direct measurements from satellites did not exist (i.e. before the eighties). Moreover, Chatzistergos *et al.* (2022) established, using historical collections of spectroheliograms, a relationship between sunspot and plage areas, which will greatly improve the TSI reconstruction since the Maunder minimum (1650-1710, the longest documented period of very low solar activity). Indeed, data concerning plages are not available before the photographic era (i.e. before 1900), contrarily to sunspots which were drawn since the seventeenth century.

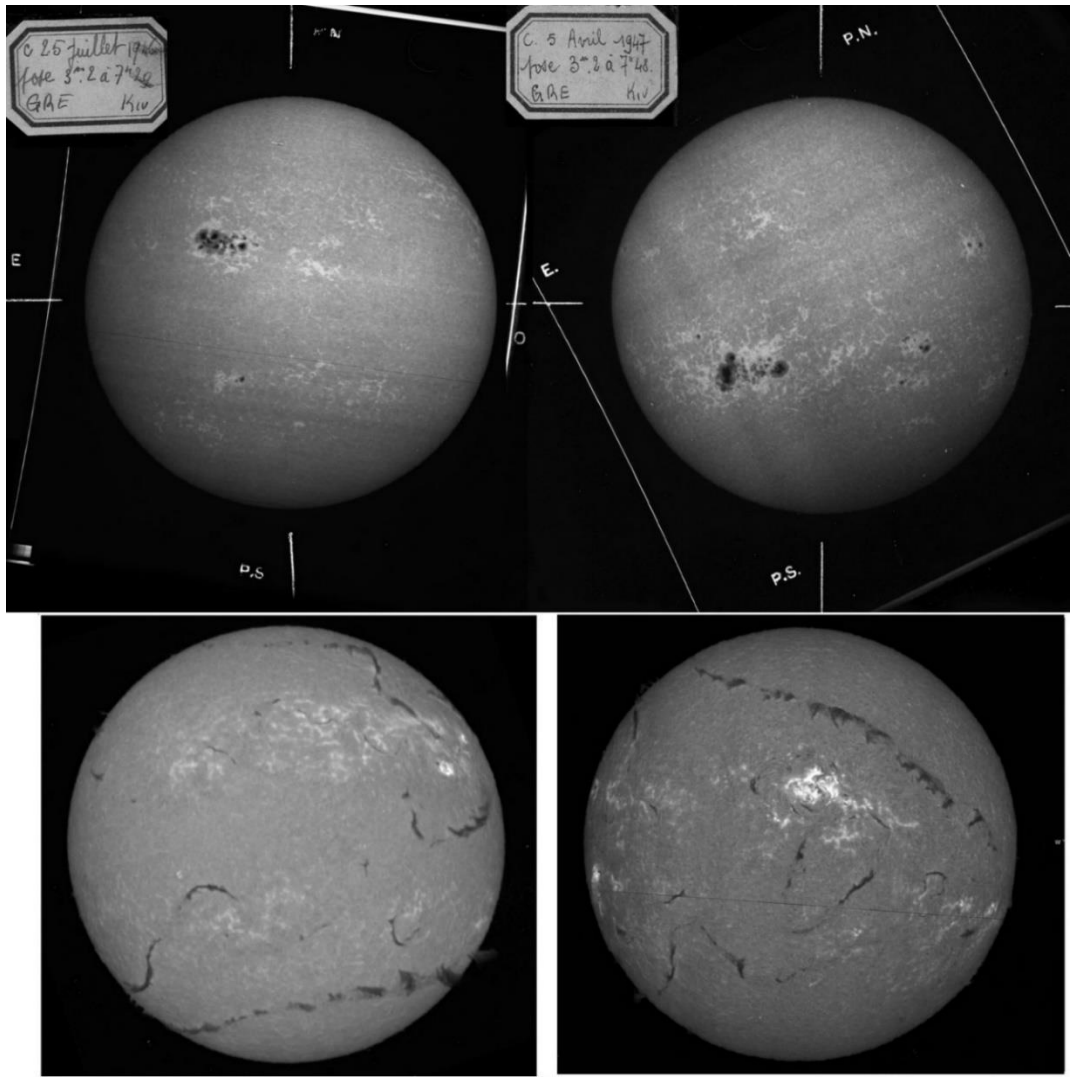

**Figure 42.** Rare events observed with the spectroheliograph. Huge sunspot groups (top, 25 July 1946 and 5 April 1947, seen in CaII K1v) and extremely long solar filaments (bottom, 30 January 1999 and 16 July 2002, observed in Hα). Courtesy Paris observatory.

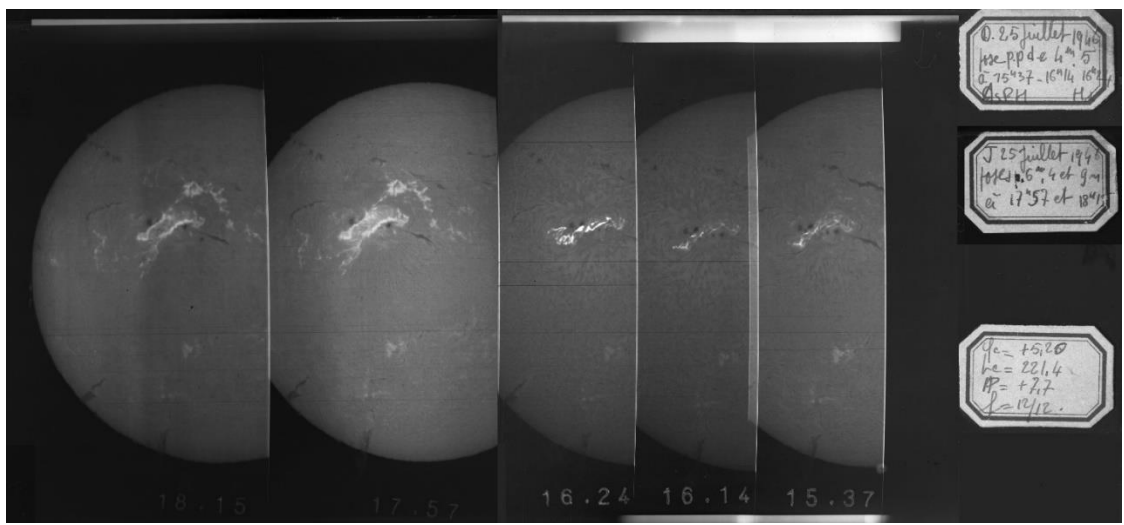

**Figure 43.** The major two-ribbon solar flare of 25 July 1946 observed in Hα with the spectroheliograph (the times are, from the right to the left, 15:37, 16:14, 16:24, 17:57, 18:15 UT). The short-duration flash phase (lasting minutes) occurs at 16:14. Two bright ribbons (17:57, 18:15) form during the long-duration gradual phase (lasting hours) and correspond to the impact of energetic particles on the chromosphere. Courtesy Paris observatory.

# 9 THE MODERN SPECTROHELIOGRAPH (1989-TODAY)

In 1980, the collection of systematic observations was rich of more than 65 000 spectroheliograms, and the two dedicated instruments (CaII K and Hα) were those of 1909, with no major modification. It was decided to change progressively all optical and mechanical parts of the imaging telescope and spectrographs, which were 70 years old, by new and higher quality elements, under the responsibility of G. Olivieri. This was a long and complicated task, because daily observations had to continue simultaneously. The main optical characteristics (such as the solar diameter and wave-bands) were preserved for backward compatibility with the existing photographic collection. New characteristics are summarized in Table 5.

In the 1989 version (Olivieri, 1989), the three prisms for CaII K and the plane grating for Hα were replaced by a single and more dispersive blazed grating with modern characteristics (17°26' blaze angle, 300 groves/mm), delivering the CaII K line in interference order 5 and Hα in order 3. The old (A2) chamber (previously for CaII K, Figure 38) was kept alive only for sporadic research purposes. The (A1) chamber was reserved for systematic observations. Photographic glass plates were abandoned, and 13 x 18 cm² films were used at the 3.0 m focus of the chamber. The bandwidth of images remained 0.15 Å and 0.25 Å, respectively for CaII K and Hα. The films were systematically digitized with a 12 bits scanner after 1995. At this date, the spectroheliograms became freely and immediately available to the solar community and to the public owing to the development (by the author of this paper) of the BASS2000 solar data base (https://bass2000.obspm.fr). The 1989 version was ready for the implementation of a digital sensor, a Reticon linear array of 2048 elements, but this device was never installed, as CCD rectangular arrays were more promising.

In 2002, the Meudon collection contained at least 85 000 spectroheliograms under the form of glass plates and films. The appearance of the electronic version (Figure 44) was a major change: the second slit (in the spectrum), the photographic films and their motorized carriers disappeared, so that the use of (A1) and (A2) chambers was abandoned. They were kept only for historical reasons. A CCD detector (a back illuminated 1340 x 100 array from Princeton Instruments, 14 bits digitization, 20 x 20 µm² pixel, running at 1 MHz) was introduced inside the spectrograph, because the focal length was reduced from 3.0 m to 0.90 m (the optical magnification decreased from 2.31 to 0.69 because of the small pixel size of the electronic sensor). There was no change for the spectral pixel (0.15 Å and 0.25 Å, respectively for CaII K and Hα), but five points were recorded simultaneously in the cores of both lines, and only one for the violet wing of the Calcium line. As the detector was not fast enough to record the full line profiles, observations of the CaII core (K2v, K3, K2r) and wing (K1v) were not simultaneous, but sequential.

The 2018 version (Figure 44) offers outstanding and state-of-the art performances (Malherbe *et al.*, 2022), with spectral pixels reduced to 0.093 Å and 0.155 Å, respectively for CaII K and Hα, and full line profiles are now registered (Figure 45). The detector is a fast and low noise scientific CMOS detector of 2048 x 2028 pixels (Fairchild CIS2020 with 6.5 x 6.5 µm² pixels, 16 bits digitization, running at 100 MHz). The focal length of the chamber is reduced to 0.40 m (magnification 0.31). As CaII H (3968 Å) and CaII K (3934 Å) form together on the chip, we record simultaneously these two violet lines. Hence, two observations are required daily, on one hand for Hα, and on the other hand, for CaII H & K. A second series is done with the disk attenuator for prominences and longer exposure time.

**Table 5.** Features of the modern versions of Meudon spectroheliograph. BW = bandwidth of monochromatic images. The focal length of the collimator is 1.30 m. The right column indicates the number of simultaneous wavelengths recorded along the line profiles (for electronic detectors only).

| Date | Chamber (m) | Hα BW (Å) | CaII K BW (Å) | Hα dispersion (Å/mm) | CaII K dispersion (Å/mm) | Wavelengths (pixels) |
|---|---|---|---|---|---|---|
| 1989 | 3.0 | 0.25 | 0.15 | 3.2 | 1.9 | 1 |
|  | slit 75 µm, FILM |  |  |  |  |  |
| 2002 | 0.9 | 0.25 | 0.15 | 12.5 | 7.5 | 5 |
|  | CCD 20 µm |  |  |  |  |  |
| 2018 | 0.4 | 0.155 | 0.093 | 28.3 | 14.3 | > 100 |
|  | sCMOS 6.5 µm |  |  |  |  |  |

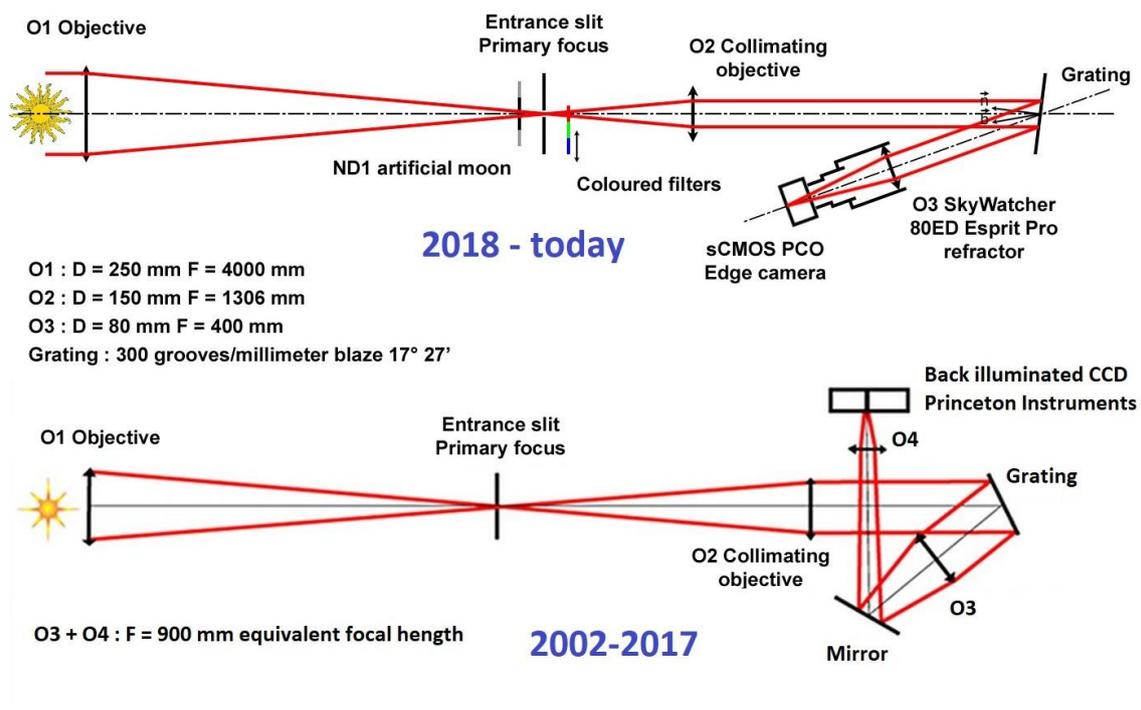

**Figure 44.** Optical designs of the spectroheliograph between 2002 and 2017 (CCD version, bottom) and after 2017 (fast sCMOS version, top). Courtesy Paris Observatory.

The spectral sampling of the successive versions of the spectroheliograph is detailed in Figure 45 and shows how the 2018 version is improved in comparison to the photographic instrument, owing to the development of fast and powerful numerical technologies. We produce today data-cubes with two space coordinates (x, y) on the Sun and one spectral coordinate (the wavelength along the line profiles). Standard observations generate 3D FITS data-cubes, from which we derive standard monochromatic images, that are slices of the cube for a given wavelength. The slices provide the backward compatibility with the photographic spectroheliograph. Data-cubes and monochromatic slices are freely available on-line at the BASS2000 database (https://bass2000.obspm.fr), in FITS format for scientific use and JPEG format for quick look, with the solar North up.

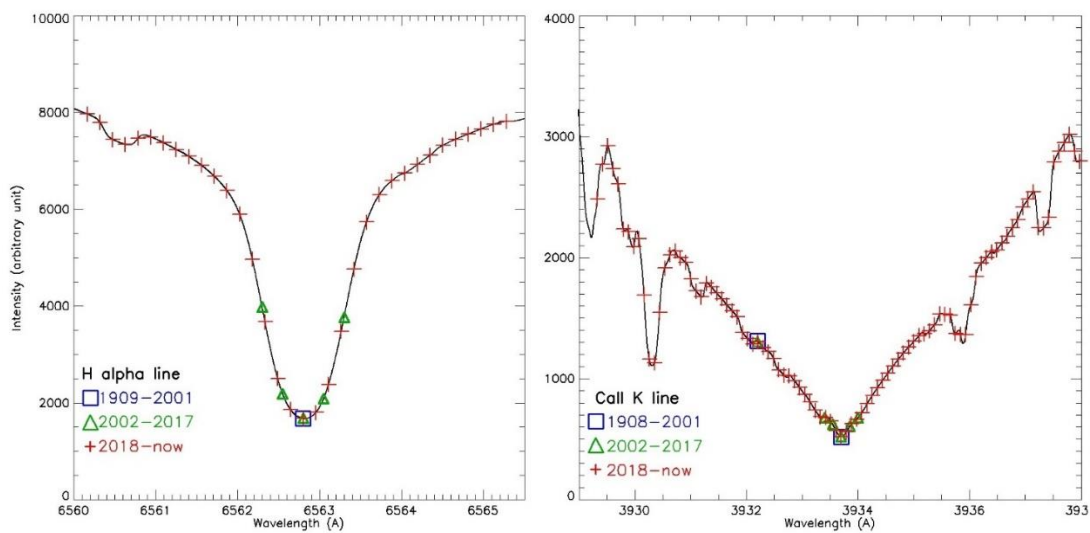

**Figure 45.** Spectral sampling of the successive versions of the spectroheliograph. Wavelength (in Angström) in abscissa. Hα (left) and CaII K (right) line profiles are displayed. Before 2001, the photographic plate provided only monochromatic images. The selected wavelengths are indicated by the black squares (Hα and CaII K line cores, plus CaII K1v blue wing). Between 2002 and 2017, five points (green triangles) were recorded at line centres of Hα and CaII K. After 2017, the full profiles are observed (red crosses) for both lines, plus CaII H (not shown). Courtesy Paris Observatory.

Concerning the photographic collection since 1908, glass plates and films (solar diameter 86 mm) after 1980 are available at BASS2000 in low resolution (250 dpi, FITS 12 bits, JPEG 8 bits, 2.25"/pixel). Images of the period 1908-1980 are only available in JPEG (https://bass2000.obspm.fr/piwigo/index.php?/category/157) under the form of monthly catalogues. However, sporadic observations since 1893 do exist. A second digitization of plates is in course, with much better resolution (600 dpi, FITS 16 bits and JPEG, 0.95"/pixel), but it is a rather slow process which requires a lot of time for operators, which are also observers. This second pass will be completed in a few years (2026) and is optimized for the average seeing (2'') of Meudon. We also have the possibility, upon request, to use a prototype scanner designed for night sky plates (4000 dpi, FITS 16 bits, 0.14"/pixel) in case of specific needs.

**CONCLUSION**

Large ground-based solar telescopes (such as the Dunn solar telescope, Sacramento Peak; the Mac Math telescope, Kitt Peak; the Vacuum Tower, GREGOR and THEMIS telescopes, Canary Islands; the giant DKIST telescope, Mauna Loa, Hawaii) and the space-borne HINODE (JAXA/NASA) have surface scanning spectrographs well suited for high spatial resolution of small fields of view (1' x 1' typical).

For the global survey of solar activity and space weather applications, observations of the full Sun at high temporal resolution (1 minute) are required. Several ground based networks (such as the seven GONG stations) made the choice of monochromatic imagery with filters, sacrificing the spectral information. Consequently, spectroheliographs are no more numerous, because of their moderate observing cadence, despite of the possibility to record full line profiles of each solar pixel. However, a space-borne compact spectroheliograph, dedicated to the Hα line, was launched on 14 October 2021 by the Chinese space administration (Chinese HAlpha Solar Explorer, CHASE, https://ssdc.nju.edu.cn). This is the first space solar mission of China (Li *et al.*, 2022) and the first satellite observing the full Sun in Hα from orbit (usually, satellites observe the million-degree corona in ultraviolet wavelengths). CHASE offers outstanding performances in terms of spatial and spectral resolutions.

The Meudon collection, spanning 10 solar cycles, remains exceptional for long-term reconstructions of solar variability and for studies of rare events. However, the future of Meudon spectroheliograph is uncertain after 2026, because of the retirements of observers which will probably not be compensated. The instrument is too complex for automation. For these reasons, we plan to include experienced amateurs in the technical team. Besides, amateurs can now build for their own compact spectroheliographs, such as the SOL'EX (http://www.astrosurf.com/solex/), with spectroscopic performances not far from the historical instruments. Modern detectors make them increasingly fast, announcing a probable regain of interest for full Sun spectroscopy in the context of solar patrol, initiated by Hale in the USA, Deslandres and d'Azambuja in France. Let us cite Martres (1996): "*All solar astronomers are inheritors of these pioneers, who organized the international survey of solar activity… Most developments and results of the last decades are based on the knowledge outlined by these researchers*".

**ON-LINE MATERIAL (ORIGINAL FIGURES, DOCUMENTS AND ADDITIONAL VIDEO CLIPS)**

High resolution figures of this paper are available here, together with some original documents cited in the references, and video clips:

https://www.lesia.obspm.fr/perso/jean-marie-malherbe/spectroheliograms/index.html

**ACKNOWLEDGEMENTS**

This paper is dedicated to the memory of M.-J. Martres and G. Olivieri, who worked respectively during five and fifteen years with M. and L. d'Azambuja, and transmitted us the knowledge of this pioneers. The author thanks C. Calderari-Froidefond for images of Paris observatory's library, I. Bualé and F. Cornu for spectroheliograms and archives of the Meudon solar collection.

**NOTES**

**[1] The solar atmosphere**. It is composed of three layers: (1) the photosphere (the visible surface, temperature decreasing from 6000 K to 4500 K in 300 km) with dark sunspots and bright faculae around, (2) the chromosphere above (temperature increasing from 4500 K to 8000 K in 2000 km) with dark filaments and bright plages (corresponding to faculae in the photosphere), (3) the ionized and hot corona (2 million K). The corona extends at million kilometres and gives rise to the solar wind (charged particles), which propagates in

the interplanetary medium. The chromosphere and the corona require spectroscopic means to reveal their structures, respectively via absorption lines of the visible spectrum or emission lines of the ultraviolet spectrum. The solar atmosphere follows a 11-year activity cycle and a 22-year magnetic cycle. Flares and coronal mass ejections occur in active regions a few years around the solar maximum; the maximum of the current cycle (number 25) is forecasted for 2025. The symbol K above (Kelvin) is the unit of the absolute temperature, related to the Celcius temperature by T(K) = T(°C) + 273.15.

**[2] The wavelength**. Electromagnetic waves, such as the visible light, are detected by the telescopes. They form a continuous spectrum (such as the rainbow) with superimposed spectral lines, revealing the presence of atoms, ions or molecules. In the visible or ultraviolet spectrum, the lines are identified by their wavelength, measured in nanometres (1 nm = $10^{-9}$ m), but spectroscopists often prefer the Angström (1 Å = 0.1 nm = $10^{-10}$ m). The visible solar spectrum (4000-7000 Å) reveals thousands of absorption lines, while emission lines occur in the ultraviolet spectrum (below 4000 Å) or at the limb.

**[3] The solar structures**. Dark sunspots are regions of intense magnetic fields (0.1-0.3 T). Bright faculae or plages form, together with sunspots, active regions, and exhibit smaller magnetic fields (0.01-0.05 T). Dark filaments, also called prominences when seen at the limb, are thin and high structures (50 000 km) of dense material suspended in the corona by weak magnetic fields (0.001 T). The symbol T is the unit of magnetic field (Tesla); the Gauss (1 G = $10^{-4}$ T) is also used. Flares occur in zones of unstable fields; reconnections convert magnetic energy into kinetic energy (ejections), radiation (X-rays) and heat (brightenings).

**[4] The Doppler effect**. When a light source moves in the observer's direction, the spectral lines are shifted towards shorter wavelengths (blueshift). On the contrary, if the source moves in the opposite direction, lines are shifted towards longer wavelengths (redshift). The Dopplershift w(Å) is a wavelength shift, proportional to the projection V of the velocity vector along the line-of-sight (it is also called radial velocity, positive towards the observer in solar physics): w = - λ (V / C), where λ is the line wavelength (Å) and C is the speed of light (3 $10^5$ km/s).

**[5] Catching the solar light**. Solar astronomers often use heliostats (optical systems with flat mirrors) to catch the solar light and reflect it in a fixed direction, which allows the use of telescopes and spectrographs located inside laboratories. The polar siderostat has a single mirror and reflects the light in the direction of the polar axis. The Foucault siderostat has also a single mirror, but the mechanics is much more complex; it reflects the solar light in the horizontal direction. Coelostats use two mirrors with simple mechanics. The primary mirror rotates around the polar axis to follow the Sun; the secondary mirror reflects the light of the first mirror in a fixed direction (either horizontal or vertical).

**[6] The Zeeman effect**. In the presence of magnetic fields, the spectral lines split in two parts (the circularly polarized signals σ+ and σ-) with respect to the position without magnetic field. The Zeeman split is w(Å) = ± 4.67 $10^{-13}$ g λ² B, where g is the Landé factor, λ the line wavelength (Å) and B (Gauss) the projection of the magnetic field vector along the line-of-sight. This effect has been discovered in laboratory by P. Zeeman in 1896 and recognized in sunspots by G. Hale in 1908. g is a quantum mechanics factor. Lines with g = 0 are insensitive to the magnetic field. Many iron lines (g in the range 2-3) allow to measure the photospheric magnetic field, and some Fraunhofer lines (such as Calcium) provide the chromospheric field above. Coronal fields are 100 times smaller than the one of sunspots or faculae, and difficult to measure; this is one of the challenges of the giant DKIST telescope in Hawaii (USA).

**[7] The Hanle effect**. The Zeeman effect is insensitive to unresolved and turbulent magnetic fields. However, anisotropic illumination of atoms creates linearly polarized light beams, and this is the case near the solar limb. In the presence of weak turbulent magnetic fields, W. Hanle discovered in 1924 that there exists a depolarization and a rotation of the polarization direction, depending on the field strength.

**THE AUTHOR**

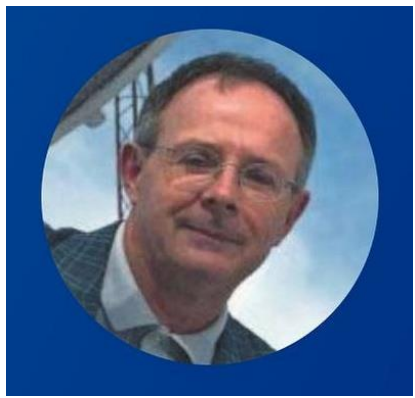

Dr Jean-Marie Malherbe, born in 1956, is astronomer at Paris-Meudon observatory. He got the degrees of "*Docteur en astrophysique*" in 1983 and "*Docteur ès Sciences*" in 1987. He first worked on solar filaments and prominences using multi-wavelength observations. He used the spectrographs of the Meudon Solar Tower, the Pic du Midi Turret Dome, the German Vacuum Tower Telescope, THEMIS (Tenerife) and developed polarimeters. He proposed models and MHD 2D numerical simulations for prominence formation, including radiative cooling and magnetic reconnection. More recently, he worked on the quiet Sun, using HINODE (JAXA), IRIS (NASA) and MHD 3D simulation results. He is responsible of the Meudon spectroheliograph since 1996.